# Anisotropic Interlayer Force Field for Group-VI Transition Metal Dichalcogenides

Wenwu Jiang,[1#] Reut Sofer,[2#] Xiang Gao,[2#] Alexandre Tkatchenko,[3,4] Leeor Kronik,[5] Wengen Ouyang,[1*] Michael Urbakh,[2*] and Oded Hod[2]

[1]*Department of Engineering Mechanics, School of Civil Engineering, Wuhan University, Wuhan, Hubei 430072, China*

[2]*School of Chemistry and The Sackler Center for Computational Molecular and Materials Science, Tel Aviv University, Tel Aviv 6997801, Israel*

[3]*Fritz-Haber-Institut der Max-Planck-Gesellschaft, Faradayweg 4-6, D-14195 Berlin, Germany*

[4]*Physics and Materials Science Research Unit, University of Luxembourg, L-1511, Luxembourg*

[5]*Department of Molecular Chemistry and Materials Science, Weizmann Institute of Science, Rehovoth 76100, Israel*

#: These authors contribute equally to this work.

[*]Corresponding authors. Email: w.g.ouyang@whu.edu.cn, urbakh@tauex.tau.ac.il

ABSTRACT

An anisotropic interlayer force field that describes the interlayer interactions in homogeneous and heterogeneous interfaces of group-VI transition metal dichalcogenides ($MX_2$ where M = Mo, W and X = S, Se) is presented. The force field is benchmarked against density functional theory calculations for bilayer systems within the Heyd−Scuseria−Ernzerhof hybrid density functional approximation, augmented by a nonlocal many-body dispersion treatment of long-range correlation. The parametrization yields good agreement with reference calculations of binding energy curves and sliding potential energy surfaces. It is found to be transferable to TMD junctions outside the training set that contain the same atom types. Calculated bulk moduli agree with most previous dispersion corrected DFT predictions, which underestimate available experimental values. Calculated phonon spectra of the various junctions under consideration demonstrate the importance of appropriately treating the anisotropic nature of layered interfaces. Considering our previous parameterization for $MoS_2$, the interlayer potential enables accurate and efficient large-scale simulations of the dynamical, tribological, and thermal transport properties of a large set of homogeneous and heterogeneous TMD interfaces.

Motivated by the fascinating physical properties of graphene, the study of alternative semiconducting two-dimensional (2D) layered materials, in particular the vast family of transition metal dichalcogenides (TMDs), has seen tremendous growth in the past decade.[1] TMDs are characterized by the general chemical formula $MX_2$, where M is a transition metal atom (e.g., Mo or W) and X is a chalcogenide atom (e.g., S, Se, or Te). The formation of moiré superlattices at heterogeneous or misaligned homogeneous TMD interfaces leads to many unique electrical,[2-6] optical,[7-9] thermal,[10-14] and tribological[15-19] properties. Notably, the most interesting physics arises at interfaces characterized by a small lattice mismatch, which is either intrinsic to the contacting materials or enforced via interfacial misalignment. This, in turn, results in relatively large moiré supercell dimensions, which are difficult to model by first-principles calculations. A viable alternative is the use of classical interlayer force-fields, which in the case of 2D materials have to include a dedicated term that accounts for the anisotropic nature of the layered construct, characterized by a covalent intralayer network and weaker dispersive interlayer interactions. When appropriately parameterized against state-of-the-art density functional theory (DFT) reference datasets, such force fields, known as interlayer potentials (ILPs), provide a desirable balance between accuracy and computational efficiency.[20-28]

Various density functional approximations that address dispersion interactions have been successfully used for generating reference data against which ILPs have been parameterized for a variety of layered material contacts.[20-28] These include the non-local (NL) vdW-DF-C09,[29] as well as DFT+D,[30, 31] Tkatchenko-Scheffler (TS),[32] many-body dispersion (MBD),[33, 34] and MBD-NL[35] dispersion augmented functionals. The obtained ILPs have proven to provide good agreement with structural,[36-39] mechanical,[27, 40-42] tribological,[25, 43-51] and thermal transport[27, 52-54] experimental observations of layered contacts, and have demonstrated predictive power that has led to novel experimental findings.[43, 44, 55] Notably, previous Kolmogorov-Crespi (KC) type anisotropic potentials have been parameterized for TMD interfaces against DFT reference data at the level of the local density approximation (LDA) augmented by the non-local vdW-DF-C09 density functional.[26] While the developed force-field provided a satisfactory description of moiré superlattice structural transformations in twisted bilayer $MoS_2$,[26] this approach is known to overestimate the binding energy of $MoS_2$, graphite, and *h*-BN by ~40%.[28, 56]

Here, we extend our anisotropic ILP parameterization to TMD interfaces consisting of tungsten (W) and selenide (Se) using MBD-NL DFT reference data that were shown to account well for long-range dispersion interactions in polarizable interfaces.[35] Taken together with our previous parameterization for molybdenum disulfide ($MoS_2$) stacks,[28] the new ILP capabilities now allow for a description of interlayer interactions in any homogeneous or heterogeneous $MX_2$ type junctions, where M = Mo, W and X = S, Se.

The reference dataset include a series of binding energy (BE) curves and sliding potential energy surfaces (PESs), calculated for homogeneous and heterogeneous TMD bilayers. The bilayer configurations have been constructed by rigidly stacking and shifting two pre-optimized monolayers. Binding energy curves for each bilayer junction have been calculated at five high symmetry stacking modes, two parallel modes (AB and AA), and three anti-parallel modes (AA', AB', and A'B), as shown in **Figure 1**. The interlayer distance has been varied in the range of 5.5-15 Å, including the sub-equilibrium regime, which is critical for the investigation of tribological properties. Each BE curve and sliding PES contains 31 and 132 data points, respectively. All calculations were performed using the FHI-AIMS code.[57] The HSE functional,[58] augmented by the MBD-NL dispersion correction,[35] has been used with the tier-2 basis-set[59] and tight convergence settings with all grid divisions and a denser outer grid. The atomic zeroth-order regular approximation (ZORA) was employed to describe relativistic effects in the vicinity of the nucleus.[57] A $k$-point grid of $19 \times 19 \times 1$ was used, with a vacuum size of 100 Å to prevent spurious interaction with neighboring image cells. Convergence tests of the calculation parameters are provided in Sec. 1 of the Supporting Information (SI). We note that the bilayer reference data are sufficient for a reliable parameterization of the interlayer interactions in the corresponding bulk systems.[28] Comparative test calculations, performed at the PBE+MBD-NL level of DFT, are presented in SI section 5, suggesting that both approaches yield similar results.

Reference data and corresponding ILP parameters for Mo-Mo, Mo-S, and S-S were inherited from our previous work.[28] To obtain the rest of the parameters, we created reference datasets for all three remaining homojunctions ($MoSe_2/MoSe_2$, $WS_2/WS_2$, and $WSe_2/WSe_2$) and two of the six possible heterojunctions ($MoS_2/MoSe_2$ and $MoS_2/WS_2$). The resulting ILP parameterization was then benchmarked against DFT calculations for the other heterojunctions, to test for accuracy and transferability of the parameters.

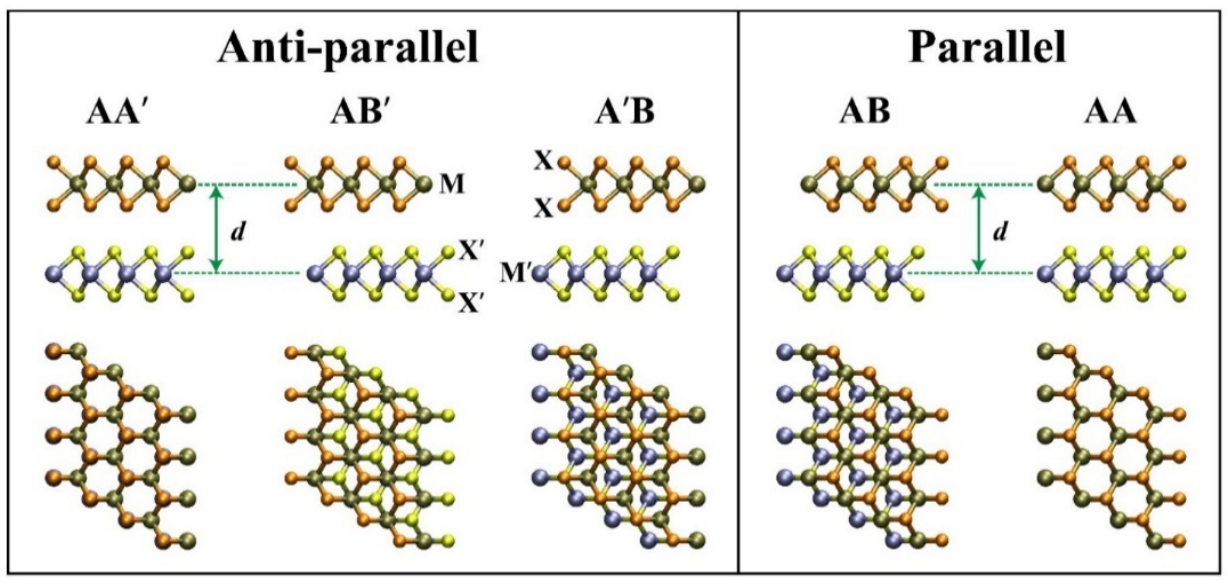

**Figure 1.** Schematic of the high symmetry stacking modes of bilayer TMD homo- and heterojunctions $MX_2/M'X'_2$ considered for BE curve calculations. For clarity of presentation, atoms residing in different layers are shown with different colors.

**Figure 2** presents calculated BE curves for the three bilayer homojunctions and the two chosen bilayer heterojunctions, at the five high-symmetry stacking modes. In general, for all bilayer junctions considered the BE curves can be divided into two groups: (i) AA and A'B stacking arrangement curves that present similar lower binding energies and larger equilibrium interlayer distances; (ii) AA', AB, and AB' stacking arrangements that show similar higher binding energies and smaller equilibrium interlayer distances. The BEs of the stacking modes appear in the following order: AA' ≥ AB > AB' > A'B ≥ AA, consistent with our previous results for $MoS_2$ and with independent PBE+D2 results.[60, 61] We note that RPA calculations[35] predict a somewhat different order, where A'B is less stable than AA. The higher stability we obtain for the AA' and AB stacking modes also agrees with the fact that they are the two natural forms ($2H$ and $3R$ phases, respectively) of TMDs, where the former is more dominant, similar to the case of hexagonal boron nitride (*h*-BN). A quantitative analysis shows that DFT results for TMD bilayer equilibrium distances and binding energies, obtained at the level of HSE+MBD-NL, differ from RPA calculations[60] by ~1.5% and 20.3% (calculated as the absolute difference between the RPA and DFT values relative to the RPA result), respectively (see **Table 1**). In comparison, vdW-DF-C09 was found to overestimate BEs by ~40%.[26, 56] **Figure 3** and **Figure 4** present corresponding sliding PESs for the homogeneous $MoSe_2$/$MoSe_2$ and the heterogeneous $MoS_2$/$WS_2$ interfaces, calculated at their fixed equilibrium interlayer distances. The left columns present the DFT reference data for the parallel (panel a, starting from the AA stacking mode) and anti-parallel (panel d, starting from the AA' stacking mode) configurations. PESs for the other homo- and heterojunctions are provided in SI Sec. 2. In total, 25 BE curves and 10 sliding PESs serve as a reference dataset in the ILP parameterization procedure.

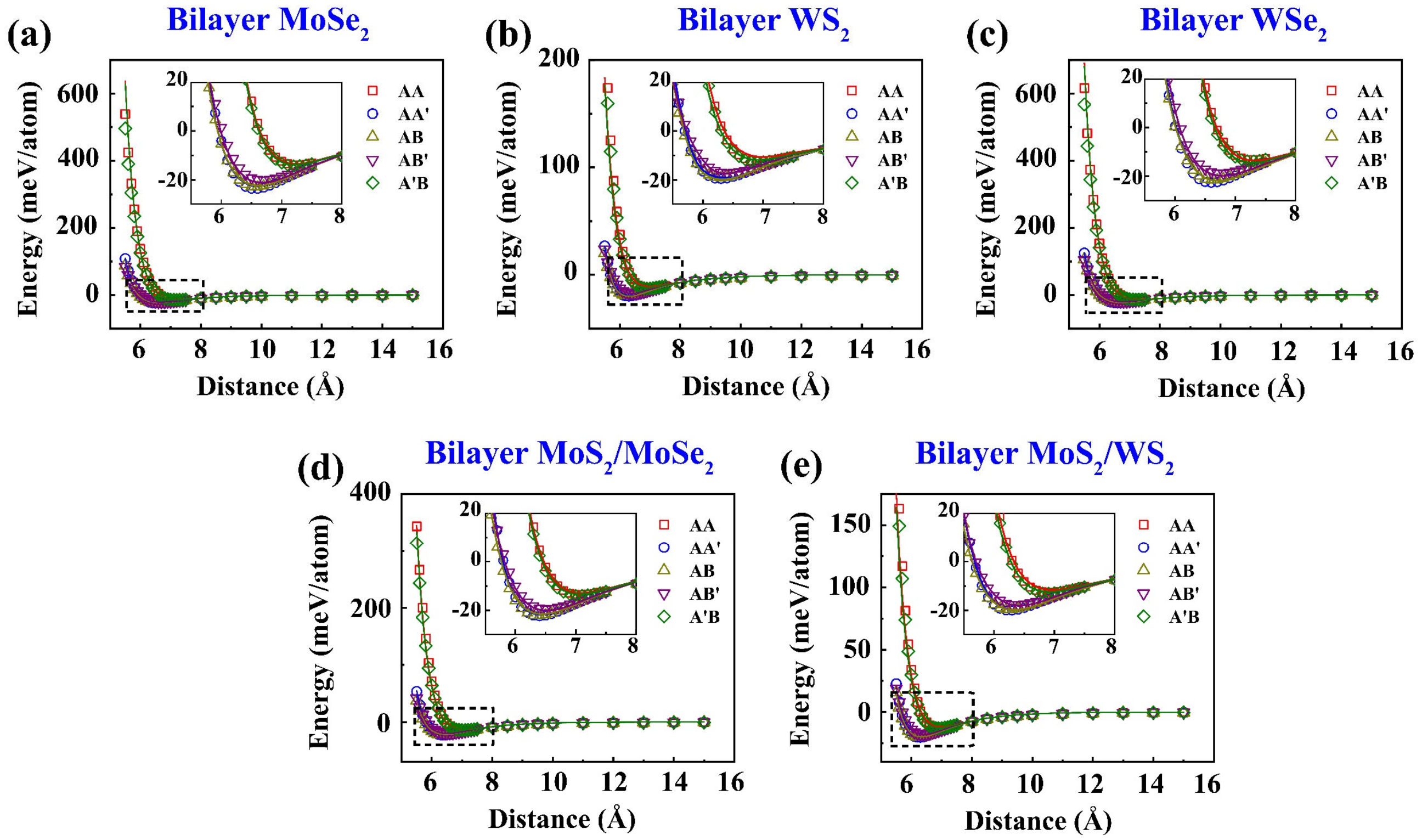

**Figure 2.** Binding energy curves for three homojunction bilayers: (a) $MoSe_2$, (b) $WS_2$, and (c) $WSe_2$, and two heterojunction bilayers: (d) $MoS_2/MoSe_2$, and (e) $MoS_2/WS_2$, calculated at the HSE+MBD-NL level of DFT (open symbols), along with the corresponding ILP fits (solid lines). Three stacking modes of the anti-parallel configuration (AA' – blue circles, AB' – purple inverted triangles, and A'B – green diamonds) and two stacking modes of the parallel configuration (AB – tan triangles, and AA – red squares) are considered (see **Figure 1**). The parameters presented in **Table S1** in SI Sec.3 are used to perform the ILP calculations. The reported energies are measured relative to the values of the two monolayers and are normalized by the total number of atoms in the unit-cell (6 atoms). The insets provide a zoom-in on the equilibrium interlayer distance region, marked by dashed black rectangles.

**Table 1.** Equilibrium interlayer distances, $d_{eq}$ (Å), and binding energies, $E_b$ (meV/atom), for $MoSe_2$, $WS_2$, and $WSe_2$ bilayers, calculated at several stacking modes using various DFT methods and the ILPs constructed in this work.[a]

| Methods | | | RPA[60] | | PBE+D2[60] | | HSE06+MBD-NL (this work) | | ILP-MBD-NL (this work) | |
|---|---|---|---|---|---|---|---|---|---|---|
| Stacking modes (Bilayer) | | | $d_{eq}$ | $E_b$ | $d_{eq}$ | $E_b$ | $d_{eq}$ | $E_b$ | $d_{eq}$ | $E_b$ |
| **$MoSe_2$** | Anti-Parallel configurations | AA' | 6.48 | 29.5 | 6.53 | 35.6 | 6.6 | 23.7 | 6.6 | 22.7 |
| | | AB' | 6.53 | 25.4 | 6.63 | 31.9 | 6.7 | 20.2 | 6.7 | 21.2 |
| | | A'B | 7.12 | 16.7 | 7.10 | 22.2 | 7.2 | 14.2 | 7.2 | 13.8 |
| | Parallel configurations | AB | 6.47 | 28.4 | 6.53 | 35.1 | 6.6 | 22.9 | 6.6 | 22.5 |
| | | AA | 7.18 | 17.2 | 7.13 | 14.4 | 7.3 | 13.8 | 7.2 | 13.4 |
| **$WS_2$** | Anti-Parallel configurations | AA' | 6.24 | 27.6 | 6.24 | 30.2 | 6.3 | 19.7 | 6.4 | 19.2 |
| | | AB' | 6.27 | 22.3 | 6.24 | 27.0 | 6.4 | 17.2 | 6.4 | 17.4 |
| | | A'B | 6.78 | 14.8 | 6.80 | 19.3 | 6.9 | 12.7 | 6.9 | 11.8 |
| | Parallel configurations | AB | 6.24 | 24.8 | 6.24 | 29.6 | 6.3 | 19.1 | 6.3 | 19.8 |
| | | AA | 6.84 | 14.6 | 6.84 | 18.9 | 6.9 | 12.4 | 7.0 | 10.6 |
| **$WSe_2$** | Anti-Parallel configurations | AA' | 6.50 | 30.3 | 6.54 | 42.0 | 6.6 | 22.6 | 6.7 | 21.8 |
| | | AB' | 6.62 | 24.5 | 6.59 | 37.3 | 6.8 | 18.8 | 6.8 | 19.8 |
| | | A'B | 7.24 | 16.1 | 7.08 | 26.3 | 7.3 | 13.6 | 7.3 | 14.4 |
| | Parallel configurations | AB | 6.54 | 28.1 | 6.54 | 40.9 | 6.6 | 21.5 | 6.7 | 21.8 |
| | | AA | 7.24 | 16.2 | 7.09 | 25.5 | 7.3 | 13.2 | 7.3 | 13.5 |

[a]Intralayer hexagonal lattice constants of 3.26, 3.16, and 3.27 Å are used for $MoSe_2$, $WS_2$, and $WSe_2$, respectively.

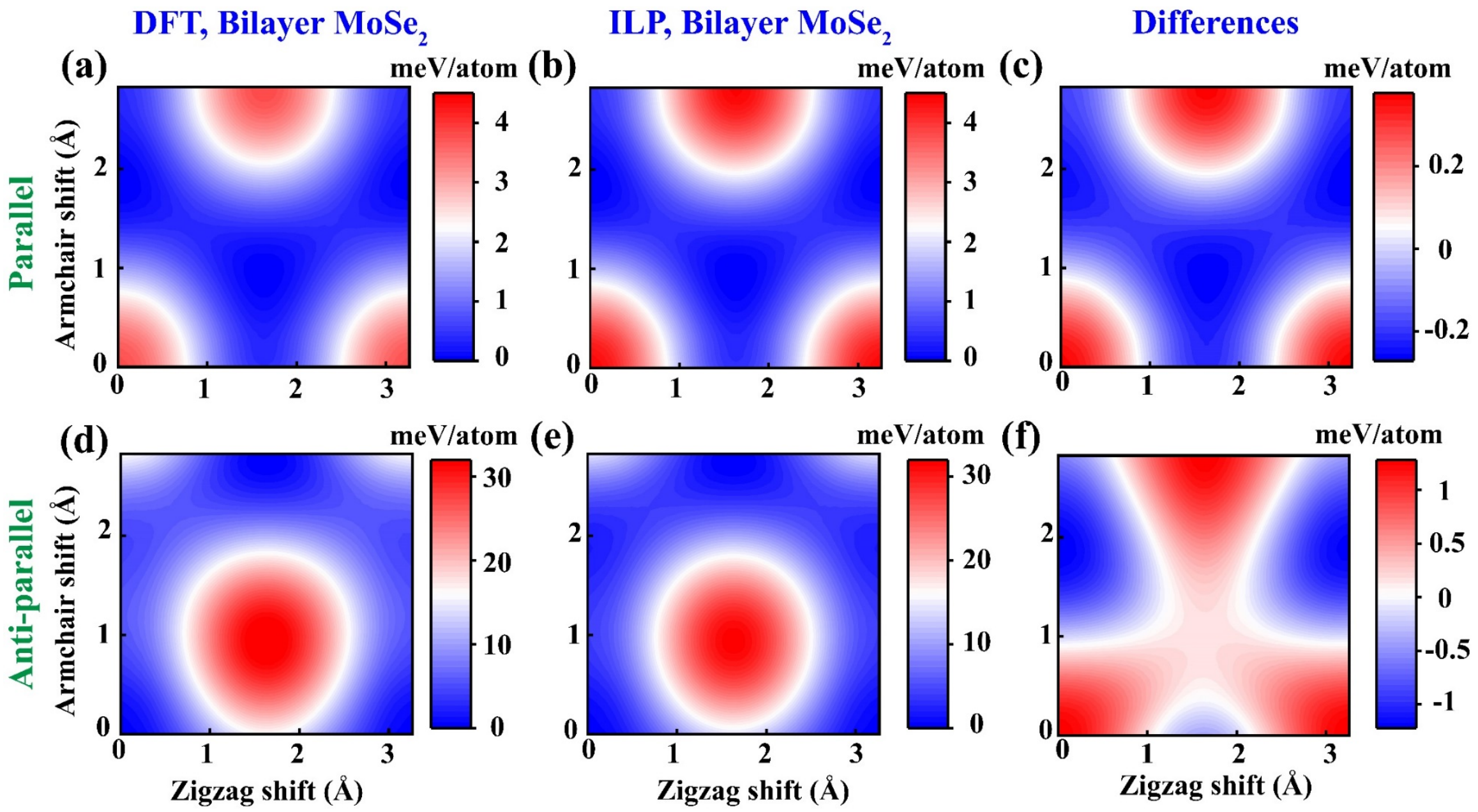

**Figure 3.** Sliding energy surfaces of bilayer $MoSe_2$, calculated at an interlayer distance of 7.2 Å for the parallel (top panels) and 6.5 Å for the antiparallel (bottom panels) configurations. The left and middle columns present DFT (HSE+MBD-NL) and ILP PESs, respectively, and the right column presents their difference maps. The parameters presented in **Table S1** in SI Sec.3 are used for the ILP calculations. The reported energies are normalized by the total number of atoms in the unit-cell (6 atoms) and measured relative to the values obtained at the AA' and AB stacking modes for the anti-parallel and parallel configurations, respectively.

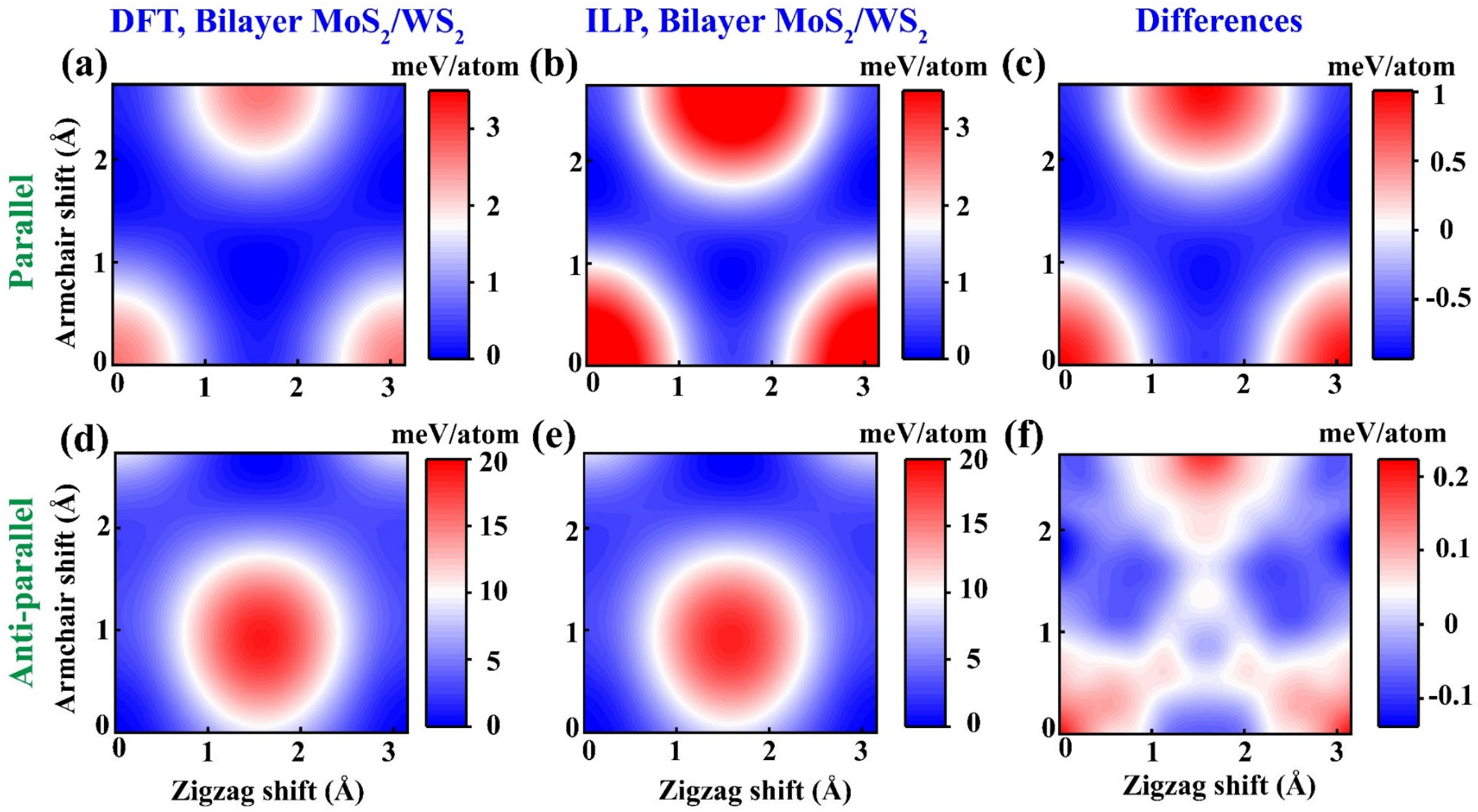

**Figure 4.** Sliding energy surfaces of bilayer $MoS_2/WS_2$, calculated at an interlayer distance of 6.9 Å for the parallel (top panels) and 6.3 Å for the antiparallel (bottom panels) configurations. The left and middle columns present DFT (HSE+MBD-NL) and ILP PESs, respectively, and the right column presents their difference maps. The parameters presented in **Table S1** in SI Sec.3 are used for the ILP calculations. The reported energies are normalized by the total number of atoms in the unit-cell (6 atoms) and measured relative to the values obtained at the AA' and AB stacking modes for the anti-parallel and parallel configurations, respectively.

The functional form of the ILP is the same as that used previously for $MoS_2$, graphene, and *h*-BN interfaces, namely:[22-25, 27, 28]

$$V(\boldsymbol{r}_{ij}, \boldsymbol{n}_i, \boldsymbol{n}_j) = \mathrm{Tap}(\boldsymbol{r}_{ij})\left[V_{\mathrm{att}}(r_{ij}) + V_{\mathrm{rep}}(\boldsymbol{r}_{ij}, \boldsymbol{n}_i, \boldsymbol{n}_j)\right], \tag{1}$$

where

$$V_{\mathrm{att}}(r_{ij}) = -\frac{1}{1+e^{-d_{ij}\left[r_{ij}/\left(s_{R,ij}\cdot r_{ij}^{\mathrm{eff}}\right)-1\right]}}\frac{C_{6,ij}}{r_{ij}^{6}} \tag{2}$$

is a long-range pairwise term that represents van der Waals attraction,

$$V_{\mathrm{rep}}(\boldsymbol{r}_{ij}, \boldsymbol{n}_i, \boldsymbol{n}_j) = e^{\alpha_{ij}\left(1-\frac{r_{ij}}{\beta_{ij}}\right)}\left\{\varepsilon_{ij} + C_{ij}\left[e^{-(\rho_{ij}/\gamma_{ij})^2} + e^{-(\rho_{ji}/\gamma_{ij})^2}\right]\right\} \tag{3}$$

is a short-range registry-dependent term that represents Pauli repulsion, and

$$\mathrm{Tap}(r_{ij}) = 20\left(\frac{r_{ij}}{R_{\mathrm{cut}}}\right)^7 - 70\left(\frac{r_{ij}}{R_{\mathrm{cut}}}\right)^6 + 84\left(\frac{r_{ij}}{R_{\mathrm{cut}}}\right)^5 - 35\left(\frac{r_{ij}}{R_{\mathrm{cut}}}\right)^4 + 1 \tag{4}$$

is a taper function that smoothly reduces the potential and its derivatives (up to the third order) to zero beyond an interatomic cut-off distance of $R_{\text{cut}} = 16$ Å. Here, $\boldsymbol{r}_{ij}$ is the interatomic distance vector between atom $i$ residing in one layer and atom $j$ residing in its adjacent layer, $\boldsymbol{n}_i$ is the surface normal vector at the position of atom $i$ calculated through its six nearest neighboring atoms residing in the same sublayer as atom $i$ (out of the three sublayers of a given TMD layer, see **Figure 5**):

$$\boldsymbol{n}_i = \frac{\boldsymbol{N}_i}{|\boldsymbol{N}_i|},\ \boldsymbol{N}_i = \frac{1}{6}\left[\sum_{k=1}^{6}\left(\boldsymbol{r}_{k,i} \times \boldsymbol{r}_{k+1,i}\right)\right], \tag{5}$$

where $\boldsymbol{r}_{k,i} = \boldsymbol{r}_k - \boldsymbol{r}_i, k = 1,2,\cdots 6$, and the summation is cyclic with $\boldsymbol{r}_{7,i} = \boldsymbol{r}_{1,i}$, $C_{6,ij}$ is the pairwise vdW attraction coefficient, $r_{ij}^{\text{eff}}$ is the sum of the effective equilibrium vdW atomic radii, $d_{ij}$ and $s_{R,ij}$ are unitless parameters determining the steepness and onset of the short-range Fermi–Dirac type damping function, $\varepsilon_{ij}$ and $C_{ij}$ are constants that define the energy scales corresponding to the isotropic and anisotropic repulsions, respectively, $\beta_{ij}$ and $\gamma_{ij}$ set the associated interaction ranges, and $\alpha_{ij}$ is a parameter that sets the steepness of the isotropic repulsion function. Importantly, the anisotropic repulsive term depends not only on the pairwise distance, $r_{ij}$, but also on the lateral interatomic distance, $\rho_{ij}(\rho_{ji})$, calculated as the distance of atom $j(i)$ to the surface normal, $\boldsymbol{n}_i(\boldsymbol{n}_j)$ of atom $i(j)$ as shown in **Figure 5**:

$$\begin{cases}\rho_{ij}^2 = r_{ij}^2 - \left(\boldsymbol{r}_{ij} \cdot \boldsymbol{n}_i\right)^2 \\ \rho_{ji}^2 = r_{ji}^2 - \left(\boldsymbol{r}_{ji} \cdot \boldsymbol{n}_j\right)^2\end{cases} \tag{6}$$

We note the interlayer Coulomb effects are found to be considerably weaker than the van der Waals interactions (see Sec. 4 of the SI), thus Coulomb effects are ignored in the ILP formula of Eq. (1), thereby considerably reducing the computational burden.[30]

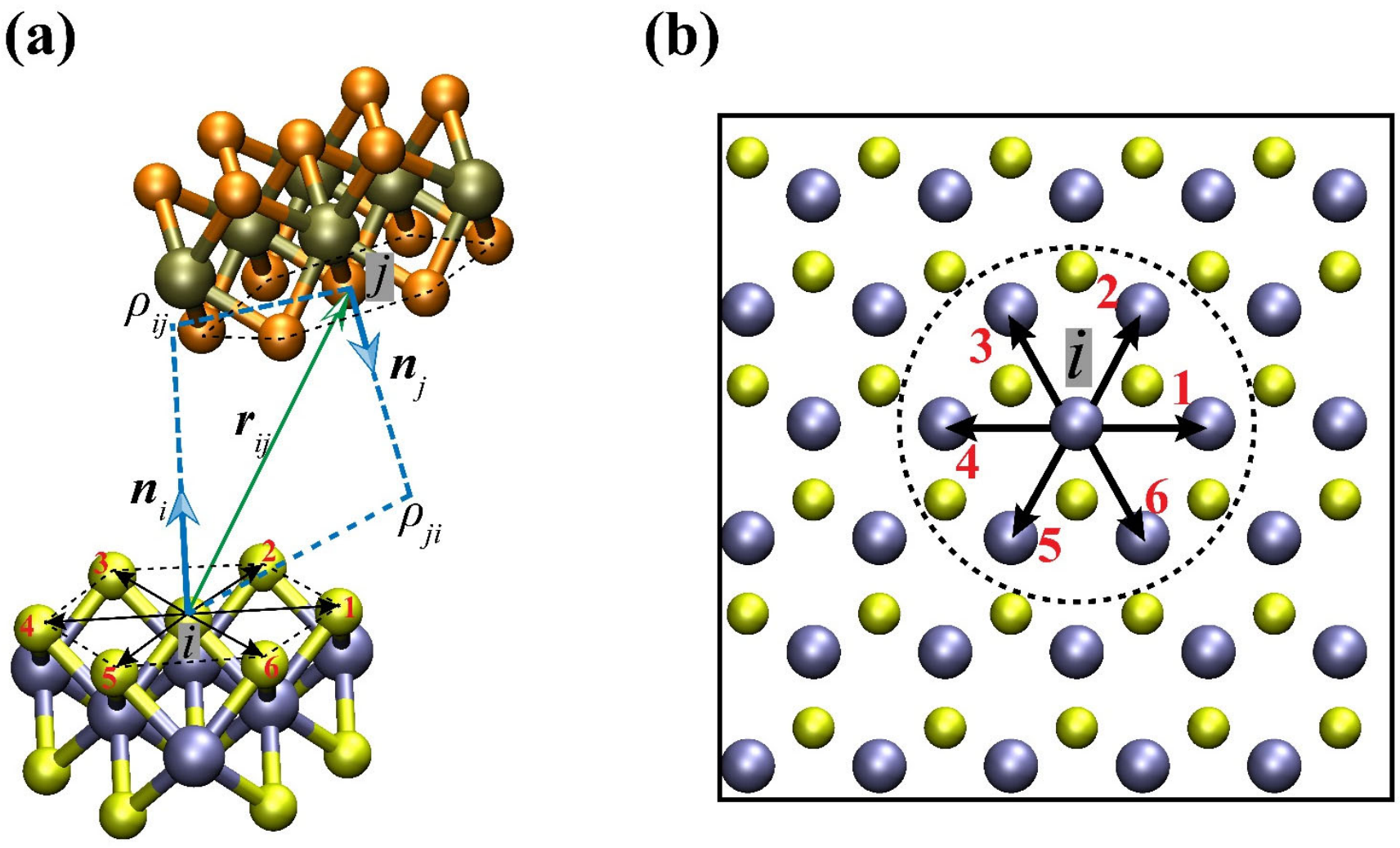

**Figure 5.** Illustration of the various distance vectors involved in the ILP expression. (a) Perspective

view of a TMD bilayer showing the full interatomic distance vector $\boldsymbol{r}_{ij}$ between atom $i$ residing in one layer and atom $j$ residing in its adjacent layer, the local normal vectors $\boldsymbol{n}_i$ and $\boldsymbol{n}_j$, and the lateral distances $\rho_{ij}$ and $\rho_{ji}$. (b) Top view of the bottom TMD layer showing the six nearest neighboring atoms of atom $i$ within the same TMD sublayer, used to define the corresponding local normal vector $\boldsymbol{n}_i$. The atomic false color scheme follows that used in **Figure 1**, where atoms residing in different layers are marked with different colors.

In total, the ILP expression involves a set of nine pairwise parameters $\boldsymbol{\xi} = \{\alpha_{ij}, \beta_{ij}, \gamma_{ij}, \varepsilon_{ij}, C_{ij}, d_{ij}, s_{R,ij}, r_{ij}^{\text{eff}}, C_{6,ij}\}$, where each parameter is symmetric with respect to index interchange due to the symmetry of the interaction. These parameters are fitted against the $M = M_{\text{b}} + M_{\text{s}}$ ($M_{\text{b}} = 25$ BE curves and $M_{\text{s}} = 10$ sliding PESs) reference DFT datasets. For convenience, we denote the BE curves as $E_m^{\text{b}}(\boldsymbol{\xi}), m \in [1, M_b = 25]$, and the sliding PESs as $E_m^{\text{s}}(\boldsymbol{\xi}), m \in [1, M_s = 10]$. To balance the large energy variation in BE curves, especially in the sub-equilibrium regime, against the relatively low energy corrugation of the PESs, we define the following objective function:

$$\Phi(\boldsymbol{\xi}) = \sum_{m=1}^{M_b} w_m^{\text{b}} \left\| E_m^{\text{b}}(\boldsymbol{\xi}) - E_m^{\text{b,DFT}} \right\|_2 + \sum_{m=1}^{M_{\text{s}}} w^{\text{s}} \left\| E_m^{\text{s}}(\boldsymbol{\xi}) - E_m^{\text{s,DFT}} \right\|_2, \quad (7)$$

where the contributions of BE curves and sliding PESs are weighted with factors of $w_m^{\text{b}}(d < d_{\text{eq}}^m) = 1$, $w_m^{\text{b}}(d \geq d_{\text{eq}}^m) = 40$ and $w^{\text{s}} = 100$, respectively, where $d$ is the interlayer distance, $d_{\text{eq}}^m$ is its equilibrium value for the $m^{\text{th}}$ BE curve, and $\|\cdots\|_2$ is the Euclidean norm (2-norm) that measures the difference between the ILP predictions and the DFT reference data. The ILP parameters are set by minimizing this objective function using an interior-point algorithm implemented in the MATLAB software suite.[62, 63] Additional details pertaining to the fitting procedure can be found in Refs. 25, 27. The fitted parameters are provided in SI Sec. 3, along with the corresponding $MoS_2$ parameters obtained in our previous study.[28] The ILP obtained using this scheme provides very good agreement with the HSE+MBD-NL reference data for both BE curves (see **Figure 2** and **Table 1**) and PESs (**Figure 3** and **Figure 4**). The maximal and average BE deviations among the five high symmetry stacking configurations depicted in **Figure 1** are 1.91 and 0.63 meV/atom, respectively. The corresponding values for the PESs are 1.42 and 0.40 meV/atom, respectively (see **Figure 3**-**Figure 4** and **Figures S3-5**).

As a test for the transferability of our ILP parameterizations, we compare the ILP BEs of the four bilayer heterojunctions not included in the reference dataset to the corresponding HSE+MBD-NL values, obtained at the high symmetry stacking modes presented in **Figure 1**. As shown in **Figure 6**, the parameterized ILP transfers well to heterojunctions outside its training set that involve the same TMD atom types, with maximal BE deviation of 2.12 meV/atom and a corresponding average value of 0.96 meV/atom. Furthermore, the transferability of the ILP is demonstrated by comparing the BE curves of AA'-stacked TMD heterojunctions that are outside the training set (see Sec. 5 of the SI).

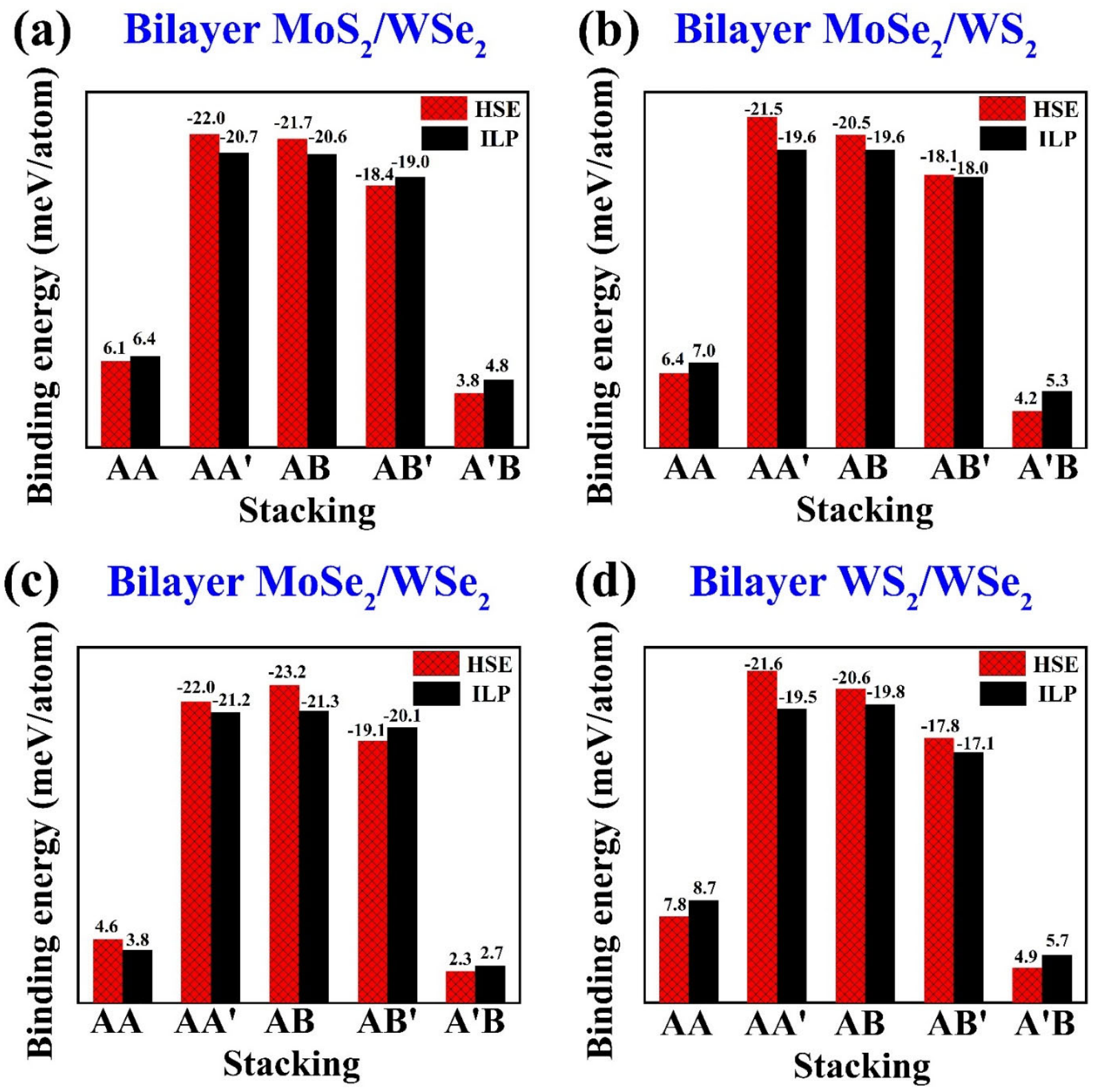

**Figure 6.** Transferability test of the developed ILP BEs, at high-symmetry stacking modes of the four bilayer heterojunctions deliberately excluded from the training set: (a) $MoS_2/WSe_2$; (b) $MoSe_2/WS_2$; (c) $MoSe_2/WSe_2$; (d) $WS_2/WSe_2$;

To further benchmark the parameterized ILP, we compare ILP-computed structural parameters ($a$ and $c$ lattice vectors and volume, $V$) of bulk $MoSe_2$, $WS_2$ and $WSe_2$, and their dependence on hydrostatic pressure ($P$), to available experimental and first-principles results. To that end, we constructed supercells consisting of twelve rectangular layers stacked in the AA' mode, each layer containing 1,250 M + 2,500 X atoms. The initial values of the $c$ lattice vector were set to 13.2, 12.8, and 13.2 Å for $MoSe_2$, $WS_2$ and $WSe_2$, respectively. The intra- and inter-layer interactions were described by the Stillinger-Weber (SW) potential and the parameterized ILP, respectively. All MD simulations were performed using the LAMMPS simulation package[64] with a time-step of 1 fs and periodic boundary conditions applied in all three directions. Nosé-Hoover thermostat with a time constant of 0.25 ps was employed to maintain the system temperature at $T$ = 300 K. Hydrostatic pressure was adjusted by relaxing the simulation box dimensions using a Nosé-Hoover barostat with a time constant of 1.0 ps.[65, 66] To obtain the *c-P, a-P,* and *V-P* curves, the simulation systems were first equilibrated in the *NPT* ensemble at a temperature of $T$ = 300 K and a fixed target pressure for 100 ps. Thereafter, structural parameters were computed by averaging over a subsequent simulation period of 100 ps. Bulk moduli were then obtained by fitting the calculated *V-P* curves (see **Figure S9**) with the Murnaghan equation of state (EOS):[67, 68]

$$V(P)/V_0 = [1 + B_V'/B_V^0 \cdot P]^{-1/B_V'}. \tag{8}$$

Here, $V_0$ and $V(P)$ are the unit-cell volumes in the absence and presence of an external hydrostatic pressure, and $B_V^0$ and $B_V'$ are the bulk modulus and its pressure derivative at zero pressure, respectively. For completeness, bulk moduli were also calculated by fitting *V-P* curves with two other commonly used equations of state: (i) the Birch-Murnaghan equation[69, 70] and (ii) the Vinet equation,[71, 72] where $B_V$ assumes a polynomial and exponential dependence on the pressure, instead of the linear dependence assumed in the Murnaghan EOS (see SI Sec. 6 for further details).

**Figure 7** presents the ILP results and available literature data for the equilibrium structural parameters $a_0$ (top row) and $c_0$ (middle row), and bulk moduli (bottom row) for the three homojunctions considered. The $a_0$ parameter, which is mostly affected by the adopted intralayer SW potential, shows good agreement with experimental results and most of the dispersion-corrected DFT values, with deviations of 0.024, 0.023, and 0.014 Å from experiments (averaged over available measured data) for bulk $MoSe_2$, $WS_2$, and $WSe_2$, respectively. Similarly, the $c_0$ parameter, which is determined by the ILP, shows good agreement with both experimental results and most of the dispersion-corrected DFT values, with deviations of 0.297, 0.419, and 0.380 Å from experiments (averaged over available measured data) for bulk $MoSe_2$, $WS_2$, and $WSe_2$, respectively. As may be expected, despite the fact that LDA predictions agree well with the experimental values, which is known to be a result of fortuitous error cancellation, dispersion corrections to DFT are important for obtaining reliable interlayer lattice constants, as demonstrated by the significantly overestimated PBE values. The Murnaghan bulk moduli of $MoSe_2$, $WS_2$, and $WSe_2$ are 39.8 ± 5.8, 38.6 ± 5.3, 34.1 ± 4.1 GPa, respectively. Similar values were obtained using the other EOSs (see **Table S2** in SI Sec. 6). These values underestimate the experimental results by ~19.5%, 32.5%, and 52.6%, respectively, and agree with most of the previous dispersion-corrected DFT results. We note that the same trends were also found for bulk $MoS_2$.

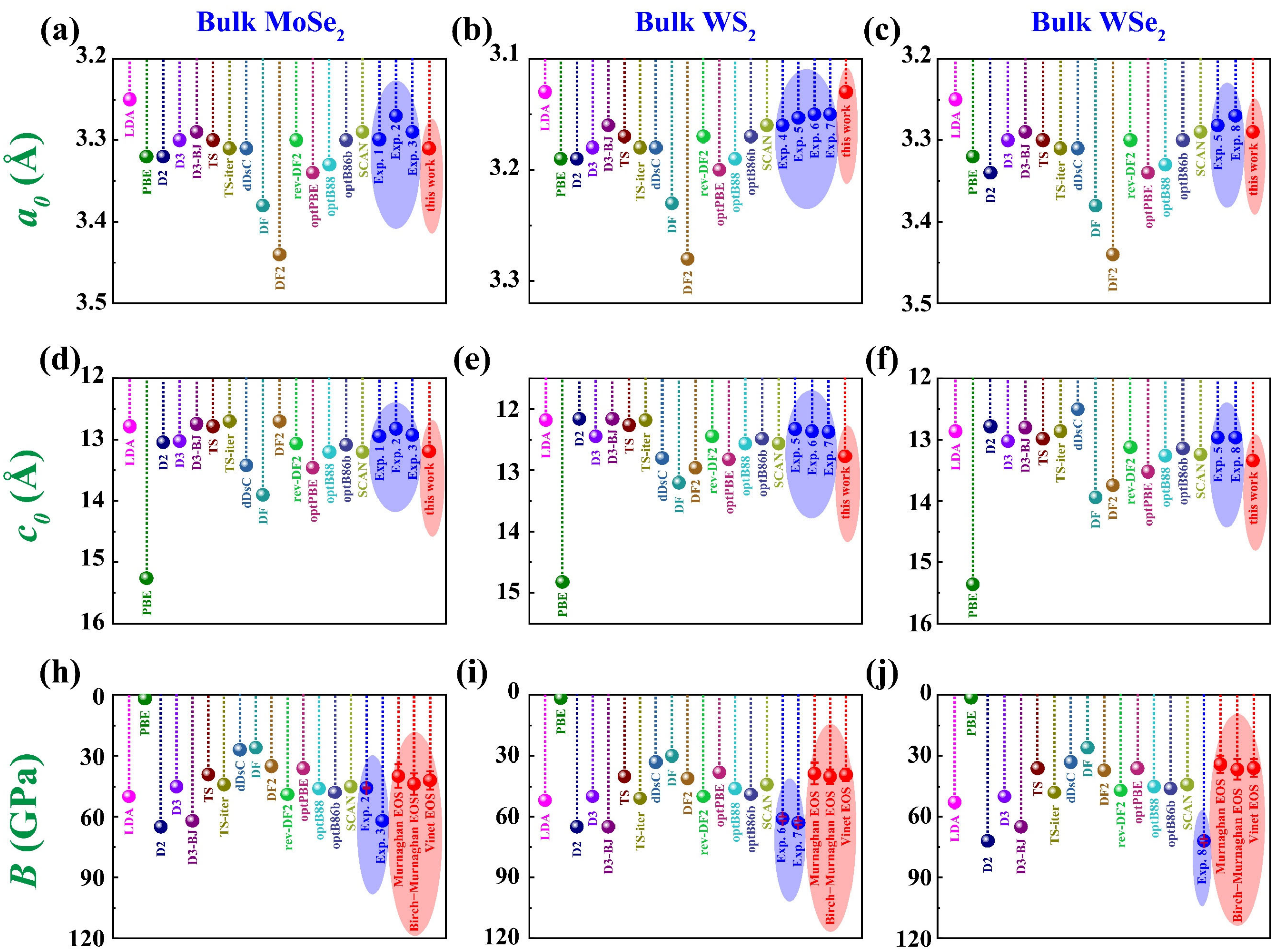

**Figure 7**. Comparison of ILP-computed $a_0$ (a)-(c), and $c_0$ (d)-(f) lattice parameters, and bulk moduli (g)-(i), with available experimental and DFT data for AA'-stacked bulk $MoSe_2$, $WS_2$, and $WSe_2$. Reported experimental values and our NPT simulation results are presented by blue and red circles, respectively. The error bars around the simulated data were obtained from the temporal standard deviation at equilibrium. In (a)-(f) they are smaller than the symbol size. The DFT reference data are taken from Ref. 73. The experimental reference data are extracted from Ref. 74 (Exp. 1), Ref. 75 (Exp. 2), Ref. 76 (Exp. 3), Ref. 77 (Exp. 4), Ref. 78 (Exp. 5), Ref. 79 (Exp. 6), Ref. 80 (Exp. 7), and Ref. 81 (Exp. 8), respectively.

Finally, we use the parameterized ILP along with the intralayer SW potential[82] to calculate the phonon spectra of the various homogeneous and heterogeneous junctions, which play a central role in their mechanical, thermal transport, and tribological properties. To that end, we follow the protocol described in Refs. 27, 28 where the dynamical matrix of the systems is first calculated in LAMMPS with a numerical differentiation step size of $10^{-6}$ Å, then diagonalized to obtain the phonon spectrum. The supercell used in these calculations contains 25×25×6 unit cells (a total of 45,000 atoms). A grid of 201 points in reciprocal space was used to plot each branch of the phonon spectrum. **Figure 8**(a) and (b) present the dispersion results for bulk $MoSe_2$ and bulk $MoS_2/WS_2$. The dispersion of the low energy out-of-plane branches (near the Γ point), which is related to the soft flexural modes of the layers and highlighted by green dotted rectangles (see **Figure 8**c-d), is mainly determined by the ILP, while the high energy modes depend mainly on the intralayer SW potential term. Notably, the isotropic Lennard-Jones (LJ) interlayer potential, parameterized by the universal force field (UFF),[83] considerably underestimates the out-of-plane phonon energies. Similar trends are found for the other bulk TMD systems considered (see SI Sec. 7).

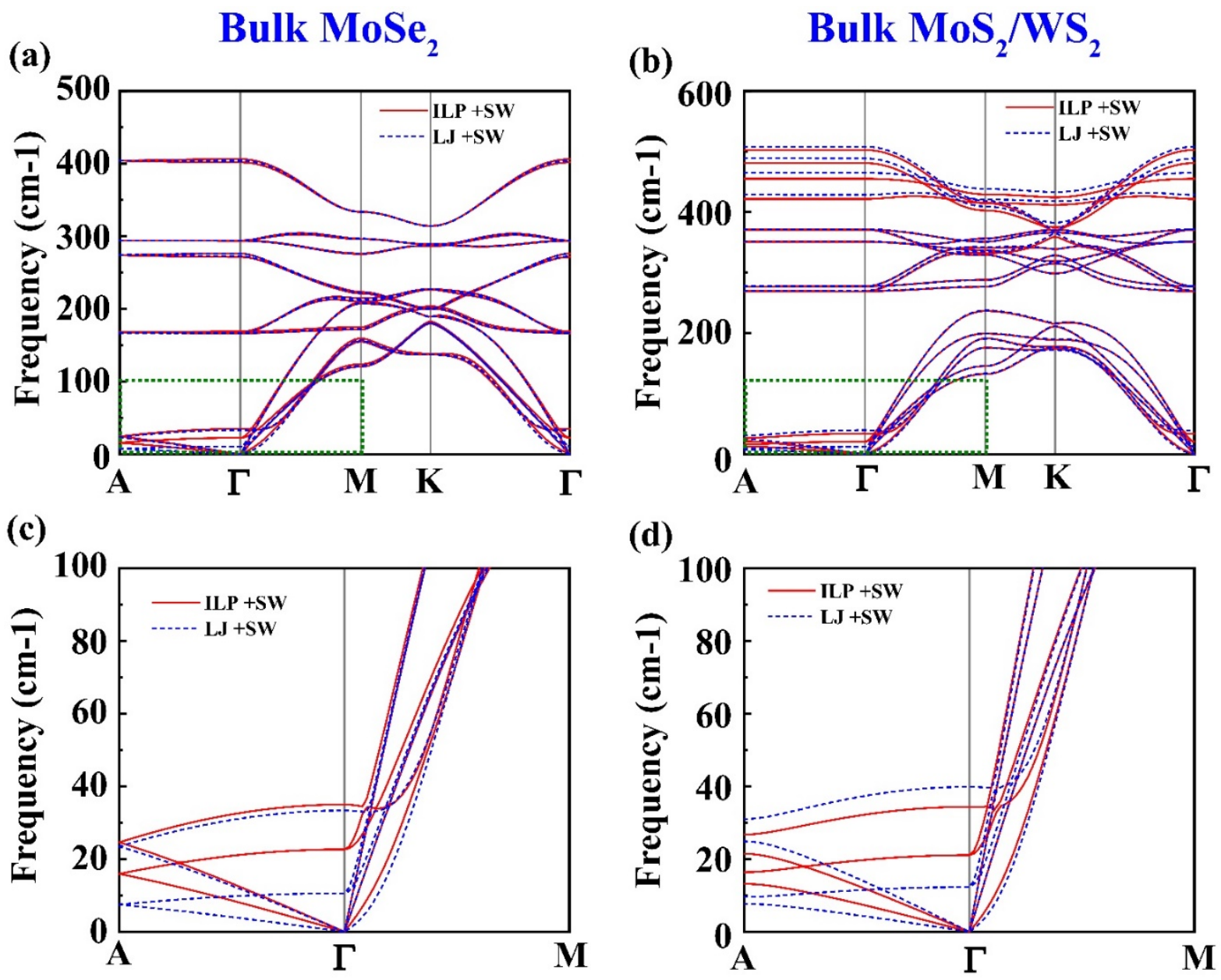

**Figure 8.** (a) Phonon spectra calculated using the ILP with the SW intralayer potential for (a) bulk $MoSe_2$ and (b) bulk $MoS_2/WS_2$. Red solid lines are dispersion curves calculated using the parameters listed in **Table S1**. LJ results are presented for comparison by the blue dashed lines. Panels (c) and (d) show zoom-ins on the low energy phonon modes near the Γ-point, marked by the dashed green rectangles in panels (a) and (b), respectively.

In conclusion, the above-presented benchmark tests demonstrate the validity range of the developed ILP for Group-VI layered TMD systems of the formula $MX_2$, where M = Mo, W and X = S, Se. The force-field parameterization against reference datasets, obtained using screened hybrid DFT augmented by nonlocal many-body dispersion corrections, yields good agreement with experimental lattice parameters and is found to be transferable to TMDs outside the training set that contain the same type of atoms. The calculated bulk moduli agree with most previous dispersion-augmented DFT predictions, which underestimate available experimental values. The successful construction of a registry-dependent interlayer potential based on state-of-the-art many-body dispersion-corrected DFT reference data for homogeneous and heterogeneous TMD interfaces opens the way for efficient yet accurate large-scale simulations of their dynamical, tribological, and heat transport properties.

ASSOCIATED CONTENT

**Supporting Information**.

Supporting Information contains the following sections: Convergence Tests of the Reference DFT Calculations; Additional Sliding Potential Energy Surfaces; Interlayer Potential Parameters; Importance of Coulomb Interactions; Transferability of the developed ILPs; Bulk Moduli Fitting; and Phonon Spectra for Additional Junctions.

AUTHOR INFORMATION

#: These authors contribute equally to this work.

**Corresponding Author**

*E-mail: w.g.ouyang@whu.edu.cn, urbakh@tauex.tau.ac.il.

**Notes**

The authors declare no competing financial interest.

## Acknowledgments

W.O. acknowledges the financial support from the National Natural Science Foundation of China (Nos. 12102307, 11890673, and 11890674), the National Natural Science Foundation of Hubei Province (2021CFB138), the Key Research and Development Program of Hubei Province (No. 2021BAA192), the Fundamental Research Funds for the Central Universities (No. 2042022kf1177), and the start-up fund of Wuhan University. X.G. acknowledges the postdoctoral fellowships of the Sackler Center for Computational Molecular and Materials Science and Ratner Center for Single Molecule Science at Tel Aviv University. M.U. acknowledges the financial support of the Israel Science Foundation, grant No. 1141/18 and the ISF-NSFC joint grant 3191/19. O.H. is grateful for the generous financial support of the Israel Science Foundation under grant no. 1586/17, Tel Aviv University Center for Nanoscience and Nanotechnology, and the Naomi Foundation for generous financial support via the 2017 Kadar Award. L.K. thanks the Aryeh and Mintzi Katzman Professorial Chair and the Helen and Martin Kimmel Award for Innovative Investigation.

# Supporting Information for "Anisotropic Interlayer Force Field for Group-VI Transition Metal Dichalcogenides"

Wenwu Jiang,[1#] Reut Sofer,[2#] Xiang Gao,[2#] Leeor Kronik,[3] Alexandre Tkatchenko,[4,5] Wengen Ouyang,[1*] Michael Urbakh,[2*] and Oded Hod[2]

[1]*Department of Engineering Mechanics, School of Civil Engineering, Wuhan University, Wuhan, Hubei 430072, China*

[2]*School of Chemistry and The Sackler Center for Computational Molecular and Materials Science, Tel Aviv University, Tel Aviv 6997801, Israel*

[3]*Department of Molecular Chemistry and Materials Science, Weizmann Institute of Science, Rehovoth 76100, Israel*

[4]*Fritz-Haber-Institut der Max-Planck-Gesellschaft, Faradayweg 4-6, D-14195 Berlin, Germany*

[5]*Physics and Materials Science Research Unit, University of Luxembourg, L-1511, Luxembourg*

#: These authors contribute equally to this work.

*Corresponding authors. Email: w.g.ouyang@whu.edu.cn, urbakh@tauex.tau.ac.il

This supporting information document includes the following sections:

## 1. Convergence Tests of the Reference DFT Calculations

Generally speaking, the convergence of the DFT reference results depends on the choice of basis-set, reciprocal-space *k*-grid density, and vacuum size. In our previous interlayer potential (ILP) parameterization for $MoS_2$,[1] we found that the tier-2 basis-set yields well converged binding energies (BEs). Therefore, we adopt this basis set also in the present study and check for BE convergence with respect to the *k*-grid density and vacuum size used in the calculations of the studied bilayer systems. **Figure S1** and**Figure S2** present the convergence of the BEs of AA stacked $MoSe_2$ and $WSe_2$ bilayers with respect to the *k*-grid density and vacuum size, respectively. In each graph only one parameter is varied, and the rest are kept the same as detailed in the main text. Red symbols mark the parameter values employed to generate the reference results used in the main text. It is seen that the choice of *k*-grid of 19×19×1 and vacuum size of 100 Å provides BE convergence to within ~0.05 meV/atom, which is satisfactory for the purposes of the present study.

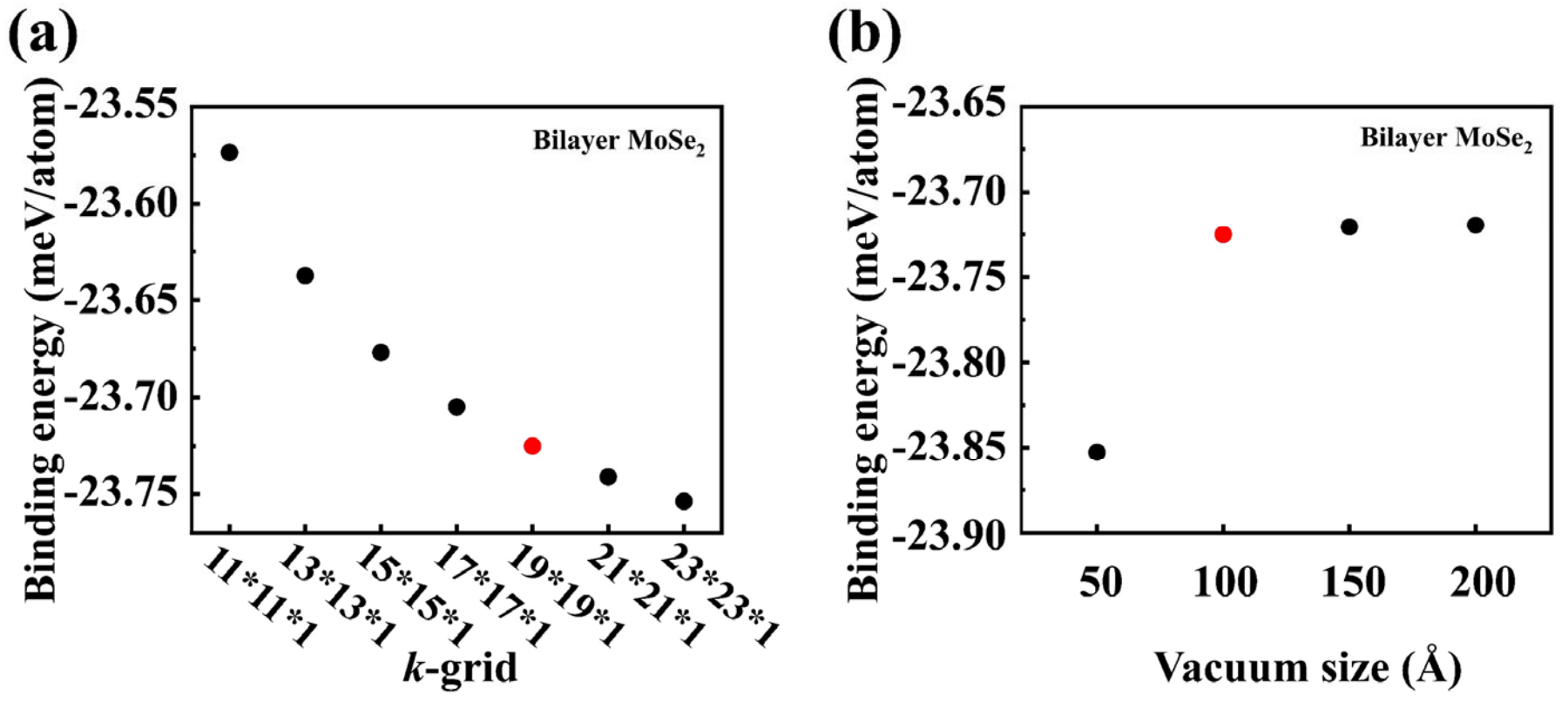

**Figure S1.** BE convergence tests for bilayer AA'-stacked $MoSe_2$ with respect to (a) the *k*-grid density and (b) the unit-cell vacuum size. The red-colored symbols mark the values used to obtain the results presented in the main text. The vacuum size used to obtain the results in panel a was fixed at 100 Å and the *k*-grid used to obtain the results presented in panel b was chosen to be 19×19×1. The interlayer distances in both panels were fixed at 6.6 Å.

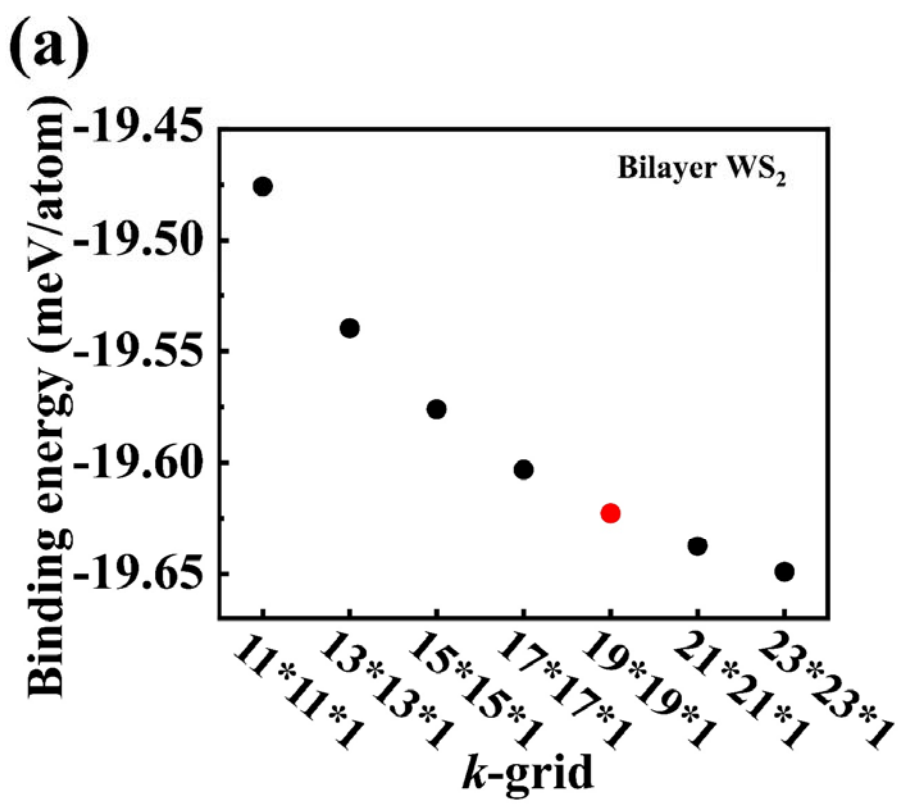

**Figure S2.** BE convergence tests for bilayer AA'-stacked $WS_2$ with respect to (a) the *k*-grid density and (b) the unit-cell vacuum size. The red colored symbols mark the values used to obtain the results presented in the main text. The vacuum size used to obtain the results in panel a was fixed at 100 Å and the *k*-grid used to obtain the results presented in panel b was chosen to be 19×19×1. The interlayer distances in both panels were fixed at 6.3 Å.

## 2. Additional Sliding Potential Energy Surfaces

In the main text, we provided sliding potential energy surfaces (PESs) for representative TMD bilayers. In this section, we provide PESs for the remaining two homojunctions (**Figure S3**: bilayer $WS_2$ and **Figure S4**: bilayer $WSe_2$) and one heterojunction (**Figure S5**: bilayer $MoS_2/MoSe_2$), which are included in the registry-dependent ILP parameterization training set. For each bilayer system, the reference DFT results (left columns), the ILP surfaces (middle columns), and their differences (right columns) are presented for the parallel (top panels) and anti-parallel (bottom panels) stacking modes. Overall, the ILP predications fit well with the DFT reference data with minor deviations of the order of ~1 meV/atom.

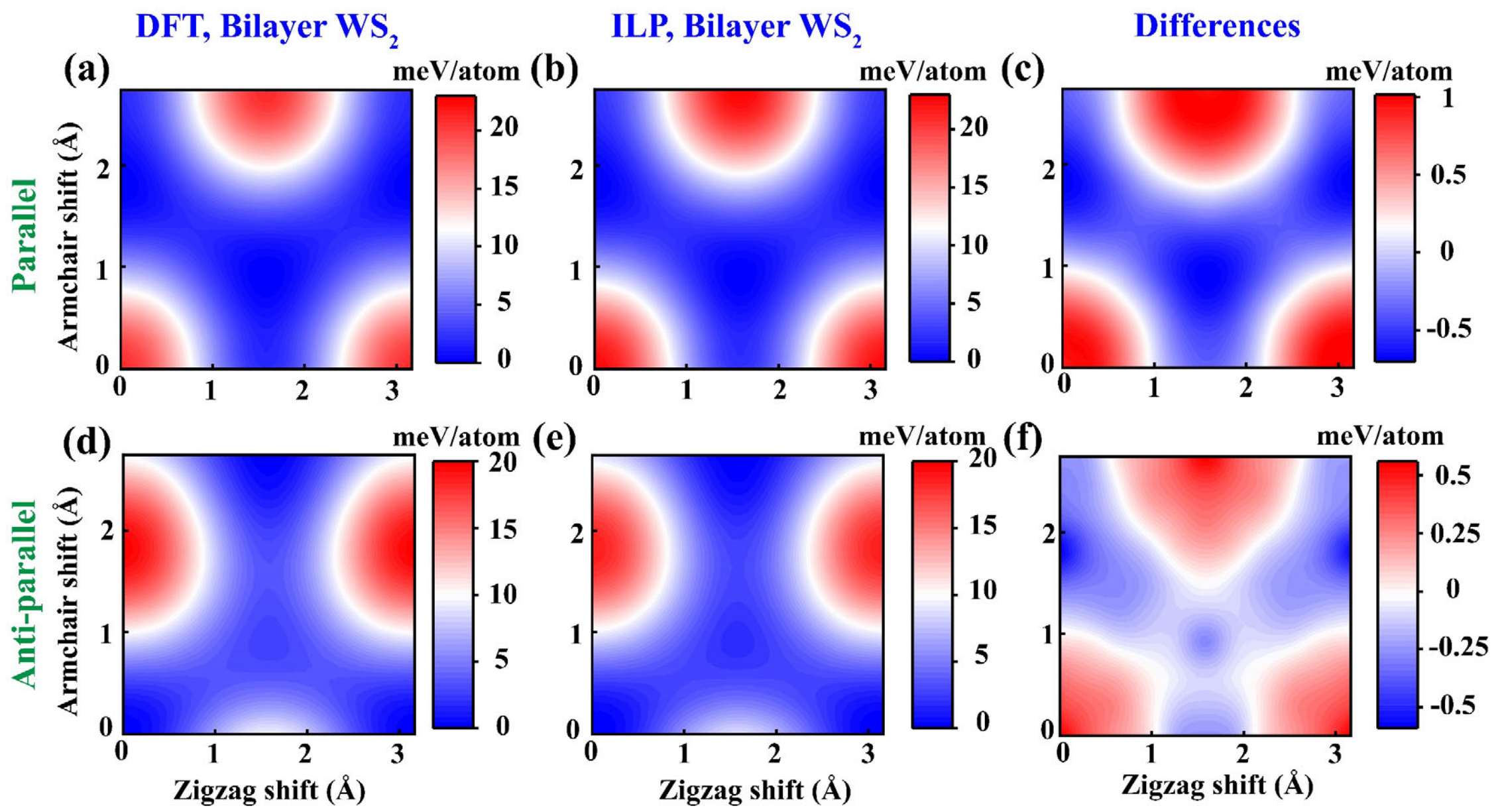

**Figure S3.** Sliding energy surfaces of the homogeneous $WS_2$ bilayer, calculated at an interlayer distance of 6.3 Å for the parallel (top panels) and antiparallel (bottom panels) configurations. The left and middle columns present the DFT (HSE+MBD-NL) and ILP PESs, and the right column presents their difference maps. The parameters presented in **Table S1** in the SI are used for the ILP calculations. The reported energies are normalized by the total number of atoms in the unit-cell (6 atoms) and measured relative to the values obtained at the AA' and AB stacking modes for the anti-parallel and parallel configurations, respectively.

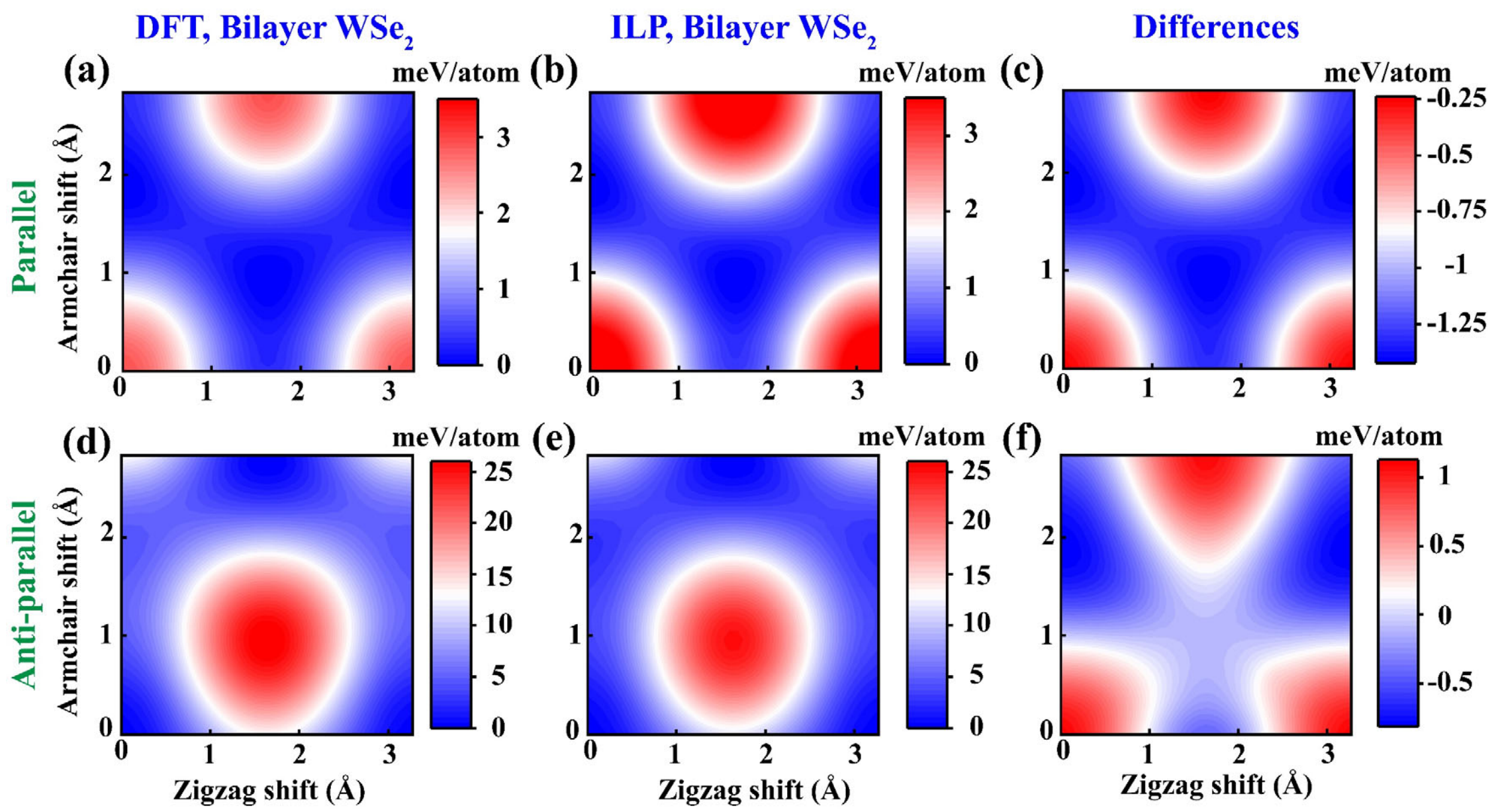

**Figure S4.** Sliding energy surfaces of the homogeneous $WSe_2$ bilayer, calculated at an interlayer distance of 7.3 Å for the parallel (top panels) and 6.6 Å for the antiparallel (bottom panels) configurations. The left and middle columns present the DFT (HSE+MBD-NL) and ILP PESs, and the right column presents their difference maps. The parameters presented in **Table S1** in the SI are used for the ILP calculations. The reported energies are normalized by the total number of atoms in the unit-cell (6 atoms) and measured relative to the values obtained at the AA' and AB stacking modes for the anti-parallel and parallel configurations, respectively.

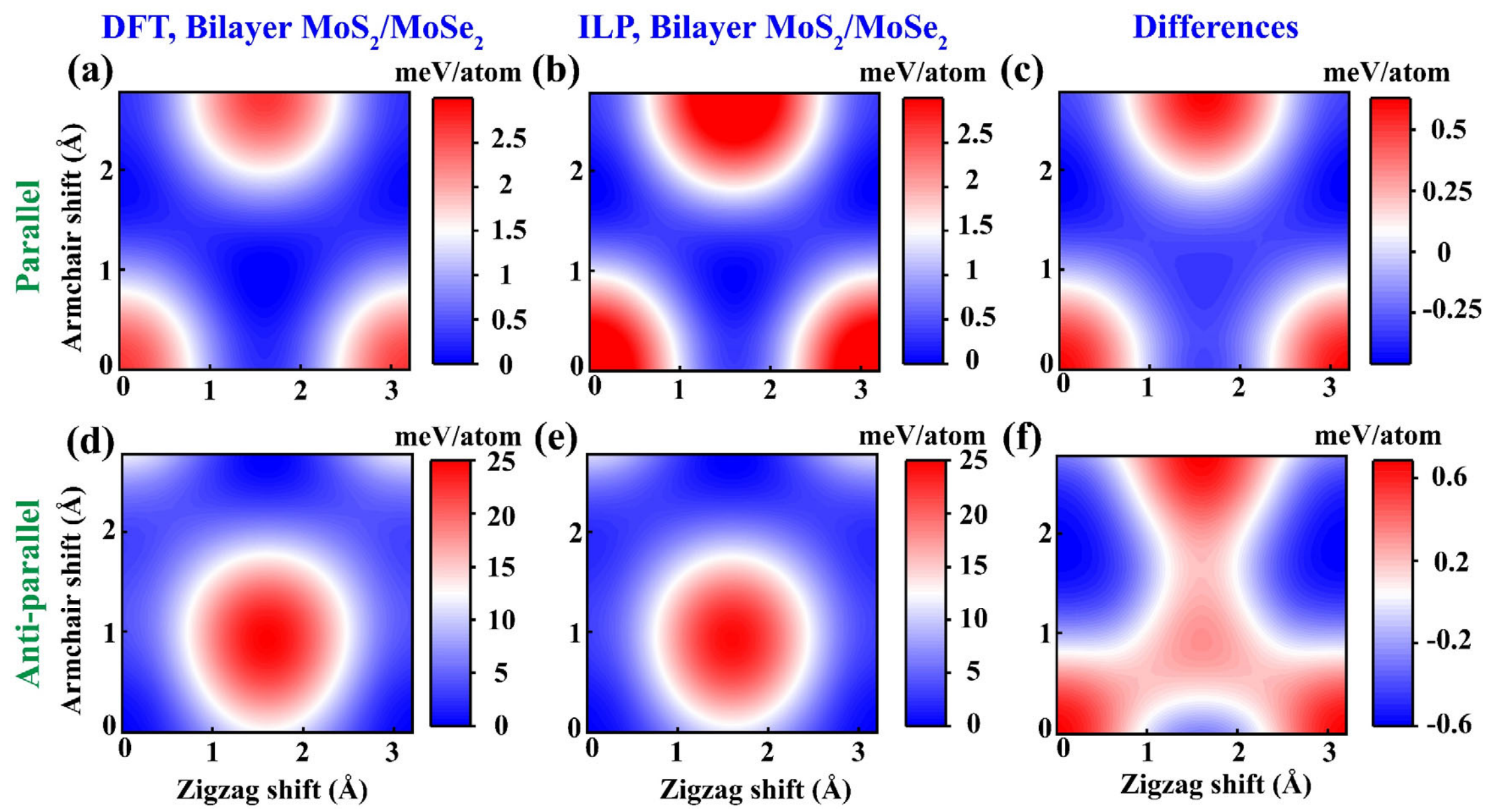

**Figure S5.** Sliding energy surfaces of the heterogeneous $MoS_2/MoSe_2$ bilayer, calculated at an interlayer distance of 7.1 Å for the parallel (top panels) and 6.4 Å for the antiparallel (bottom panels) configurations. The left and middle columns present the DFT (HSE+MBD-NL) and ILP PESs, and the right column presents their difference maps. The parameters presented in **Table S1** in the SI are used for the ILP calculations. The reported energies are normalized by the total number of atoms in the unit-cell (6 atoms) and measured relative to the values obtained at the AA' and AB stacking modes for the anti-parallel and parallel configurations, respectively.

## 3. Interlayer Potential Parameters

The ILP parameter values obtained by the fitting procedure described in the main text against DFT reference datasets at the level of HSE+MBD-NL for the various interfaces considered are presented in **Table S1**.

**Table S1**. List of ILP parameter values for laterally periodic bilayer $MX_2$ (M = Mo, W; X = S, Se). The training set includes all HSE+MBD-NL bilayer binding energy curves and sliding potential energy surfaces appearing in **Figures 2-4** of the main text and **Figure S3-Figure S5** in the SI. A value of $R_{\mathrm{cut}} = 16$ Å is used throughout. Note that all parameters are symmetric with respect to indices interchange, e.g. $\alpha_{\mathrm{Mo,S}} = \alpha_{\mathrm{S,Mo}}$.

| | $\beta_{ij}$ (Å) | $\alpha_{ij}$ | $\gamma_{ij}$ (Å) | $\varepsilon_{ij}$ (meV) | $C_{ij}$ (meV) | $d_{ij}$ | $s_{R,ij}$ | $r_{eff,ij}$ (Å) | $C_{6,ij}$ (eV·Å$^6$) |
|---|---|---|---|---|---|---|---|---|---|
| **Mo-Mo** | 5.5795 | 9.3777 | 2.0272 | 144.1518 | 97.9786 | 89.4376 | 2.0590 | 5.1221 | 491.8503 |
| **W-W** | 5.5309 | 6.6250 | 1.9832 | 0.2718 | 140.1741 | 107.3926 | 1.3563 | 4.4376 | 691.8502 |
| **S-S** | 3.1614 | 8.0933 | 1.9531 | 4.5868 | 118.0655 | 58.8094 | 0.2154 | 4.2996 | 148.8112 |
| **Se-Se** | 3.9386 | 10.5159 | 2.4158 | 3.0126 | 22.4006 | 116.8645 | 0.1511 | 5.8842 | 112.5062 |
| **Mo-S** | 3.6272 | 19.9714 | 7.5850 | 76.1019 | 3.3175 | 45.7203 | 0.9475 | 4.4104 | 150.5979 |
| **Mo-Se** | 6.1964 | 4.8441 | 14.3620 | 7.4072 | 0.0588 | 27.1562 | 0.9768 | 3.9792 | 786.0298 |
| **W-S** | 3.6801 | 11.1630 | 32.2541 | 110.0197 | 79.3813 | 138.3404 | 0.9007 | 8.8758 | 250.6008 |
| **W-Se** | 3.9420 | 21.3275 | 0.0009 | 38.7174 | 49.6776 | 1.6196 | 2.5486 | 2.8992 | 654.2895 |
| **Mo-W** | 5.4123 | 8.6471 | 2.1087 | 51.1780 | 184.3429 | 201.2813 | 2.5477 | 2.4923 | 99.9969 |
| **S-Se** | 2.8201 | 7.4912 | 1.9333s | 141.5326 | 293.1278 | 90.4709 | 0.3905 | 4.1709 | 117.6890 |

## 4. Importance of Coulomb Interactions

In the main text, we have demonstrated that the parameterized ILP, without explicit consideration of Coulomb interactions, reproduces well the binding energy curves and sliding potential energy surfaces of DFT reference data for polarizable TMD materials. This indicates that the contribution of interlayer Coulomb interactions to the binding physics should be rather weak. To confirm this, we evaluated the interlayer monopolar electrostatic interactions based on a partial-charge analysis of the different TMD atoms.

Consider, for example, bilayer $MoS_2$, where a Mulliken-charge DFT analysis, performed for the minimal unit cell at the HSE+MBD-NL level of theory, yields partial atomic charges of +0.52$e$ and -0.26$e$ for Mo and S atoms, respectively. These values can then be used within the Coulomb electrostatic energy term $E = k\frac{q_i q_j}{r}, (r < r_c)$, where $k = 14.399645\ \mathrm{eV} \cdot \mathrm{Å} \cdot \mathrm{C}^{-2}$ is the Coulomb constant, $q_i$ and $q_j$ are the effective charges of atoms $i$ and $j$ (residing in different layers), and $r_c = 16\ \mathrm{Å}$ is the cutoff. To this end, we construct a supercell consisting of a rectangular bilayer section, where each layer contains 4,900 Mo + 9,800 S atoms, within the LAMMPS package[2] and performed laterally periodic electrostatic calculations using the Ewald summation technique.

**Figure S6** and **Figure S7** present the calculated ILP BE curves and sliding PESs,[1] respectively, for the $MoS_2$ bilayer compared to the Coulomb term contribution. Notably, the latter contribution is a mere ~0.7% and ~2% of the full values of the BE and sliding potential energy barrier, respectively. This supports our choice to neglect monopolar electrostatic contributions in the ILP expression, in order to increase computational efficiency.[3]

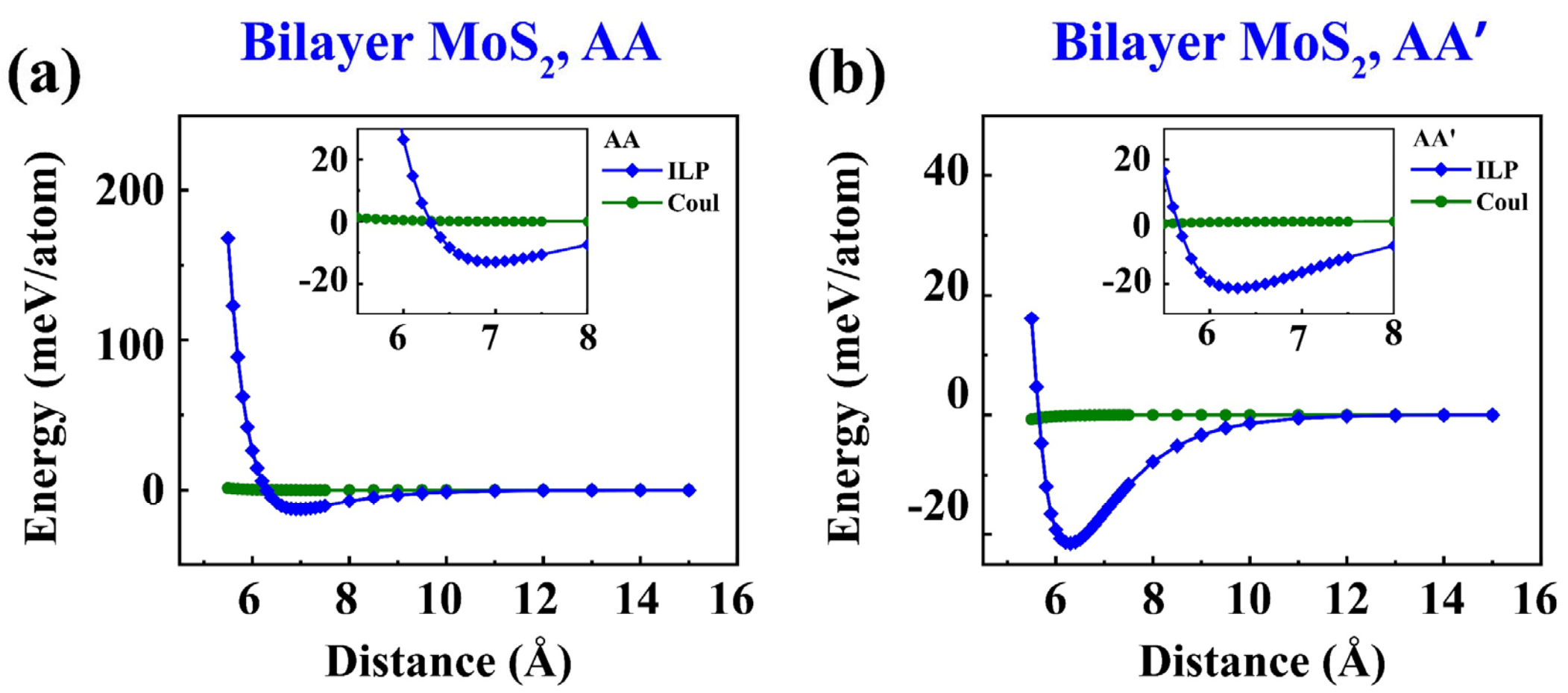

**Figure S6.** Binding energy curves of AA (a) and AA' (b) stacked $MoS_2$ bilayer calculated using the ILP (blue diamond) with the parameters of **Table S1**. The monopolar electrostatic term contribution is represented by the green circles. The reported energies are measured relative to the values of the two single layers (zero for the electrostatic term) and are normalized by the total number of atoms in the unit-cell (29,400 atoms). The insets provide a zoom-in on the equilibrium interlayer distance region.

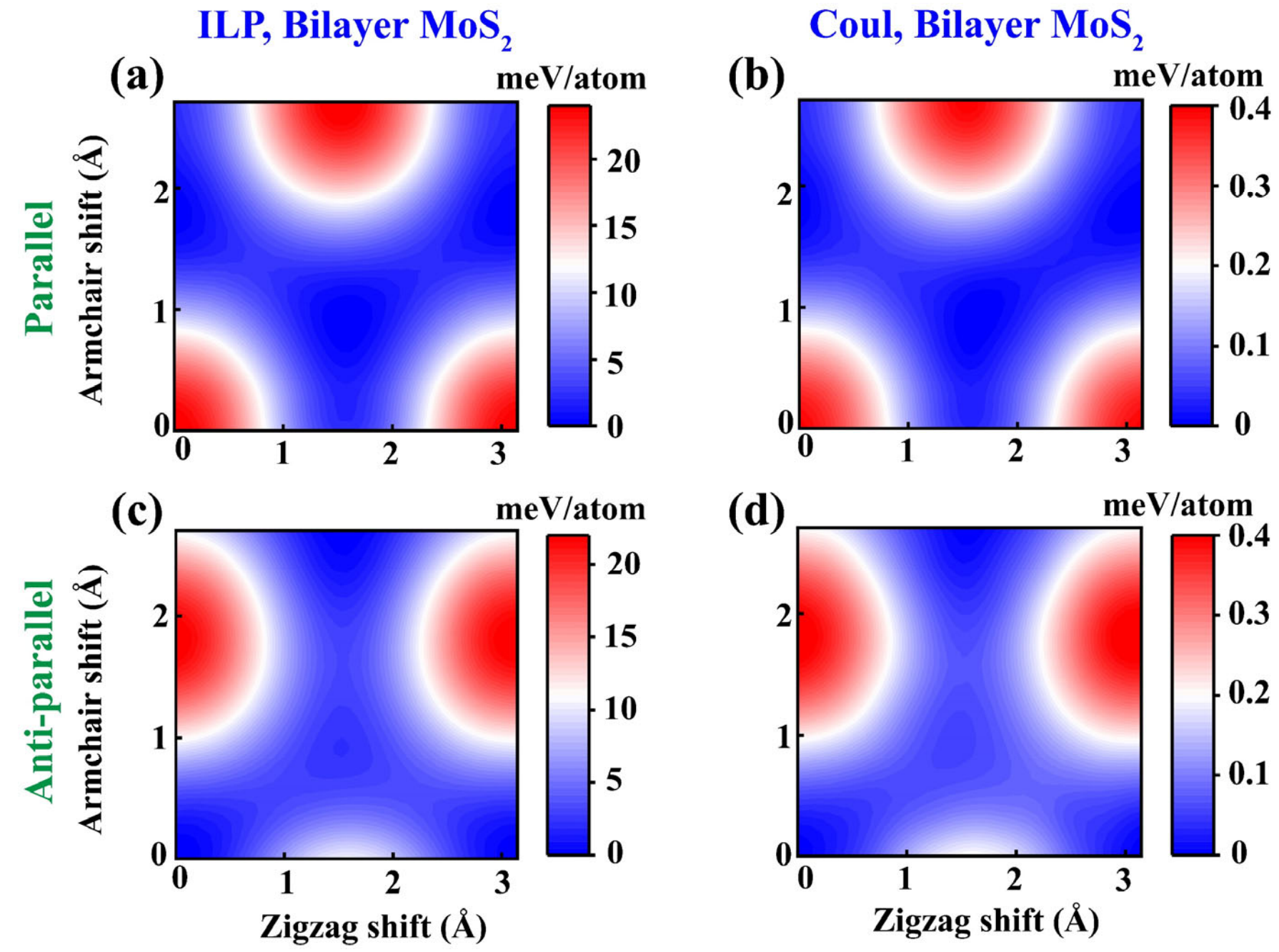

**Figure S7.** Sliding potential energy surfaces (PESs) of the bilayer $MoS_2$ homojunction, calculated at an interlayer distance of 6.24 Å for the parallel (top panels, starting from AA-stacking mode) and antiparallel (bottom panels, starting from AA′-stacking mode) configurations. ILP PESs obtained using the parameters presented in **Table S1** are presented in the left panels and the monopolar electrostatic contribution are presented in the right panels. The reported energies are normalized by the total number of atoms in the unit-cell (600 atoms) and measured relative to the values obtained at the AA' and AB stacking modes for the anti-parallel and parallel configurations, respectively.

## 5. Transferability of the developed ILPs

In this section, we demonstrate that the ILP parameterization presented in **Table S1** applies also to TMD heterojunctions that are outside the training set, as long as they contain the same atom types, which in our case, include the following AA' stacked heterojunctions: $MoS_2/WSe_2$, $MoSe_2/WS_2$, $MoSe_2/WSe_2$, and $WS_2/WSe_2$. In **Figure S8** we compare the ILP BE curves (red lines) against their DFT counterparts, calculated at the HSE+MBD-NL level of theory (black circles). The ILP BE curves compare well with the DFT calculations, with differences that are lower than ~2 meV/atom (10%) at the equilibrium separations of the different systems considered, thus demonstrating the transferability of the newly parameterized ILP.

We note that PBE+MBD-NL DFT calculations (bule triangle), which are less computationally cumbersome, provide a reasonable agreement with the HSE+MBD-NL values with differences lower than ~3 meV/atom (15%).

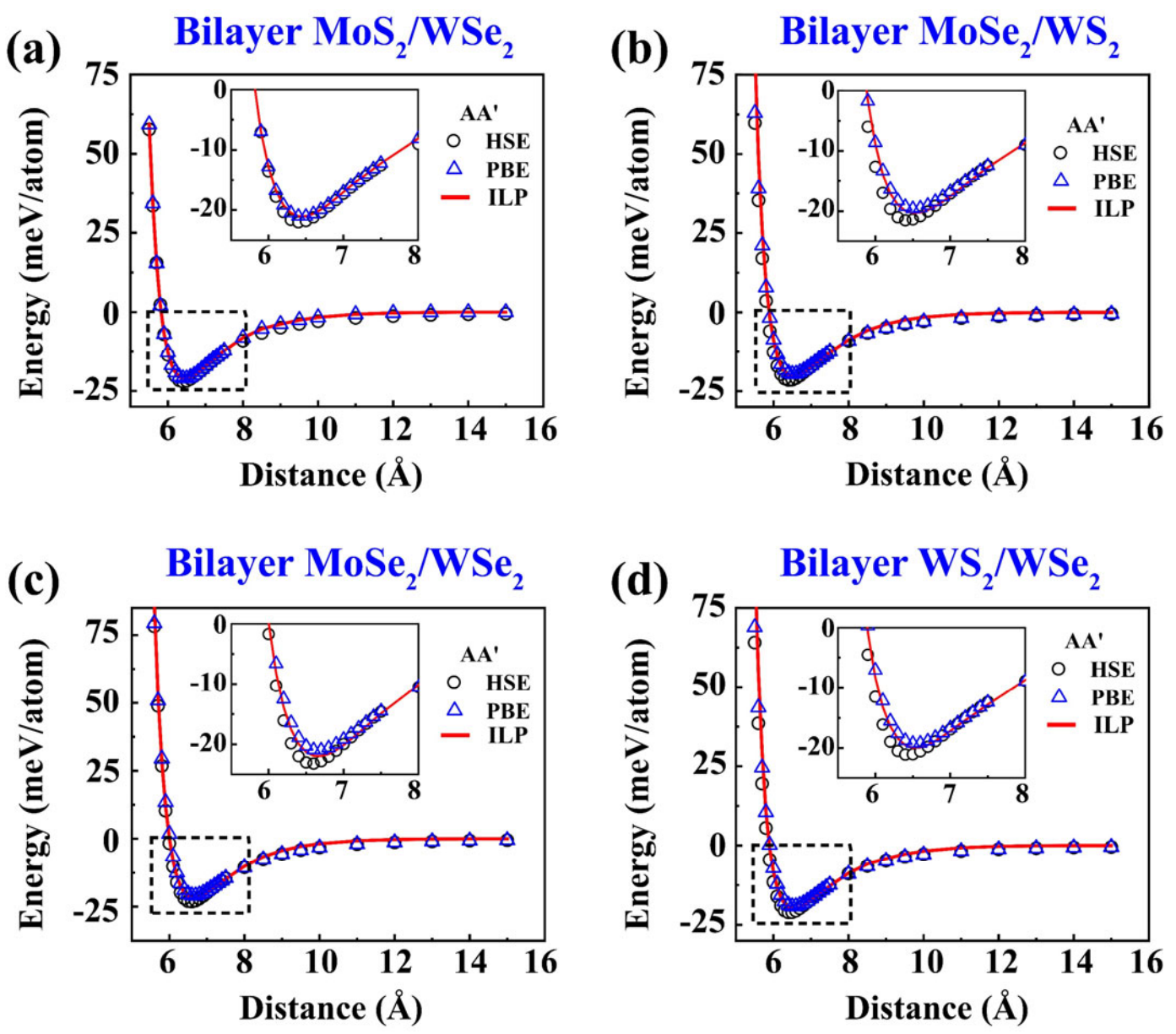

**Figure S8.** Binding energy curves for the four AA' stacked bilayer heterojunctions outside the ILP training set: (a) $MoS_2/WSe_2$, (b) $MoSe_2/WS_2$, (c) $MoSe_2/WSe_2$, and (d) $WS_2/WSe_2$ calculated using the newly parameterized (**Table S1**) ILP (red lines), and the HSE+MBD-NL (black circles) and PBE+MBD-NL (blue triangles) levels of DFT. The reported energies are measured relative to the values of the two single layers and are normalized by the total number of atoms in the unit-cell (6 atoms). The insets provide a zoom-in on the equilibrium interlayer distance region, marked by dashed black rectangles.

## 6. Bulk Moduli Fitting

The ILP pressure-volume (*P*-*V*) curves, used to evaluate the bulk moduli values reported in the main text, are presented in **Figure S9**. The evaluation is performed by fitting the ILP results with the Murnaghan equation (Eq. 8 of the main text, full blue lines),[4, 5] and two other equations of state (EOS): (i) The Birch-Murnaghan equation (S1S1S1S1S1S1S1Eq. S1, dashed red lines)[6, 7] and (ii) The Vinet equation (Eq. S2, dashed dotted purple lines),[8, 9] which take the following forms:

$$P = 3B_V^0\xi(1+2\xi)^{5/2}\left[1-\frac{3}{2}(4-B_V')\xi\right],\ \xi=\frac{1}{2}\left[\left(\frac{V}{V_0}\right)^{-\frac{2}{3}}-1\right] \tag{S1}$$

$$P = 3B_V^0\frac{(1-X)}{X^2}\exp\left[\frac{3}{2}(B_V'-1)(1-X)\right],\ X=\left(\frac{V}{V_0}\right)^{\frac{1}{3}} \tag{S2}$$

As in the Murnaghan equation, these two EOS also assume that $B_V$ varies with pressure (hence the inclusion of the derivative $B_V'$). Nonetheless, they differ in their description of the dependence of $B_V$ on the pressure, by assuming that it is linear, polynomial, and exponential for the Murnaghan, Birch–Murnaghan, and Vinet EOS, respectively.

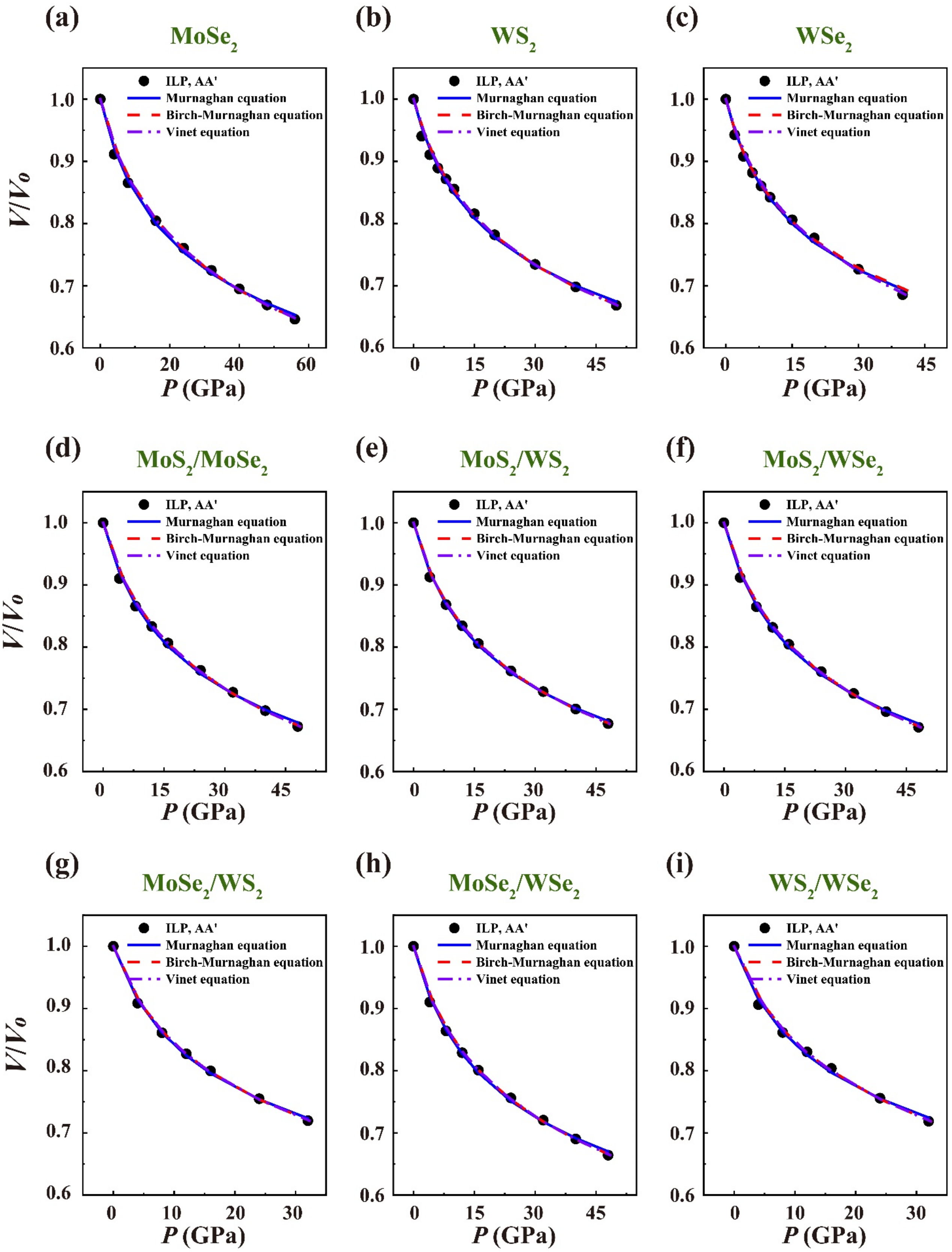

**Figure S9.** Pressure dependence of the normalized volume $V/V_0$ of nine homo- and heterojunctions. The solid black points are the NPT ILP simulations results. The solid blue, dashed red, and dashed-dotted purple lines are optimally-fitted curves obtained using Eq. 4 of the main text, Eq. S1, and Eq. S2, respectively.

The fitted bulk moduli ($B$) and its zero-pressure derivative ($B'$) for the six AA' stacked alternating heterojunctions considered are reported in **Table S2**.

**Table S2.** Bulk moduli, $B$ (GPa), and their zero-pressure derivative, $B'$ (GPa), of AA' stacked bulk $MoSe_2$, $WS_2$, and $WSe_2$, and alternating bulk heterojunctions of $MoS_2/MoSe_2$, $MoS_2/WS_2$, $MoS_2/WSe_2$, $MoSe_2/WS_2$, $MoSe_2/WSe_2$, $WS_2/WSe_2$.

| Method/Material | | **$MoSe_2$** | **$WS_2$** | **$WSe_2$** |
|---|---|---|---|---|
| Murnaghan EOS | $B$ | 39.8 ± 5.8 | 38.6 ± 5.3 | 34.1 ± 4.1 |
| | $B'$ | 4.8 ± 0.7 | 5.2 ± 0.8 | 5.5 ± 0.8 |
| Birch−Murnaghan EOS | $B$ | 43.7 ± 5.3 | 40.1 ± 3.6 | 36.6 ± 4.9 |
| | $B'$ | 4.9 ± 0.7 | 5.7 ± 0.6 | 5.8 ± 1.0 |
| Vinet EOS | $B$ | 41.9 ± 4.5 | 39.2 ± 3.1 | 35.9 ± 4.2 |
| | $B'$ | 5.5 ± 0.6 | 6.1 ± 0.5 | 6.1 ± 0.8 |
| Method/Material | | **$MoS_2/MoSe_2$** | **$MoS_2/WS_2$** | **$MoS_2/WSe_2$** |
| Murnaghan EOS | $B$ | 38.4 ± 5.1 | 38.6 ± 3.6 | 39.2 ± 4.2 |
| | $B'$ | 5.2 ± 0.7 | 5.3 ± 0.5 | 5.2 ± 0.6 |
| Birch−Murnaghan EOS | $B$ | 40.5 ± 4.9 | 38.4 ± 2.8 | 39.5 ± 4.2 |
| | $B'$ | 5.6 ± 0.8 | 6.1 ± 0.5 | 5.7 ± 0.7 |
| Vinet EOS | $B$ | 39.4 ± 4.1 | 37.6 ± 2.2 | 38.5 ± 3.5 |
| | $B'$ | 6.0 ± 0.6 | 6.4 ± 0.4 | 6.1 ± 0.5 |

| Method/Material | | **$MoSe_2/WS_2$** | **$MoSe_2/WSe_2$** | **$WS_2/WSe_2$** |
|---|---|---|---|---|
| Murnaghan EOS | *B* | 33.8 ± 4.1 | 38.2 ± 4.3 | 34.3 ± 6.8 |
| | *B′* | 5.8 ± 0.8 | 4.8 ± 0.6 | 5.7 ± 1.3 |
| Birch−Murnaghan EOS | *B* | 33.7 ± 4.0 | 39.8 ± 4.2 | 36.2 ± 6.8 |
| | *B′* | 6.7 ± 1.1 | 5.3 ± 0.7 | 6.1 ± 1.6 |
| Vinet EOS | *B* | 33.6 ± 3.3 | 38.5 ± 3.5 | 35.8 ± 5.9 |
| | *B′* | 6.8 ± 0.7 | 5.8 ± 0.5 | 6.4 ± 1.2 |

## 7. Phonon Spectra for Additional Junctions

In this section, we provide additional phonon spectra not presented in the main text. **Figure S10** presents the phonon spectra of several homogeneous and alternating heterogeneous bulk materials, as calculated using both ILP and Lennard-Jones (LJ) potentials, where the universal force field (UFF) parameters are adopted for the latter.[10] Notably, the isotropic LJ potential yields considerably lower out-of-plane phonon energies compared to those obtained using the ILP. **Figure S11** presents the phonon dispersion curves of the three bilayer homojunctions and the six bilayer heterojunctions calculated using the ILP and the LJ potential with the Stillinger-Weber (SW) intralayer potential.[11]

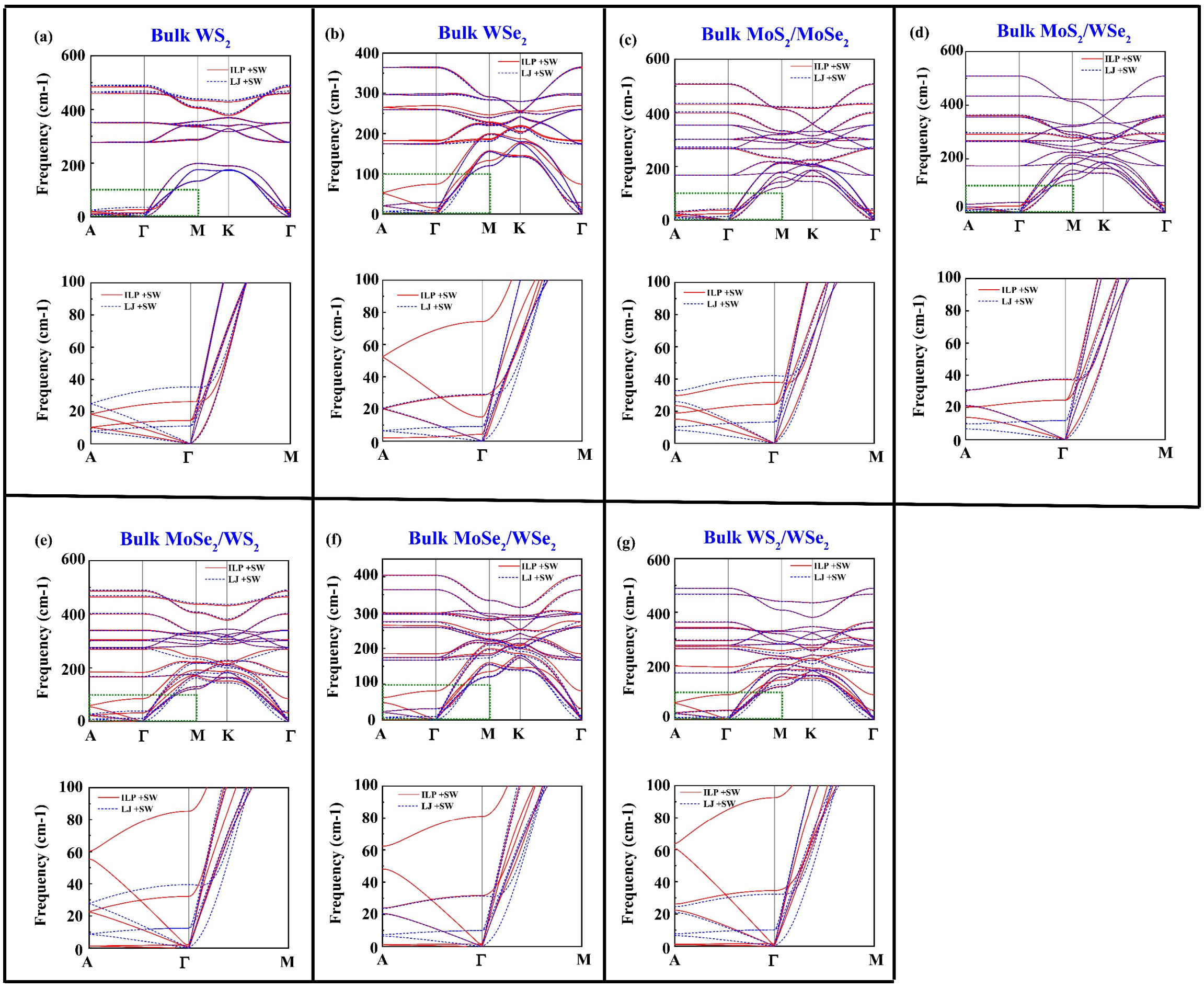

**Figure S10.** Phonon spectra calculated using ILP (red lines) and LJ (dashed blue lines) potentials with the SW intralayer potential[11] for additional bulk homo- (a,b) and heterojunctions (c-g). The green rectangles (around the Γ-point) in the top subpanels denote a zoomed-in area, shown in the lower subpanels, focusing on the low-energy phonon modes.

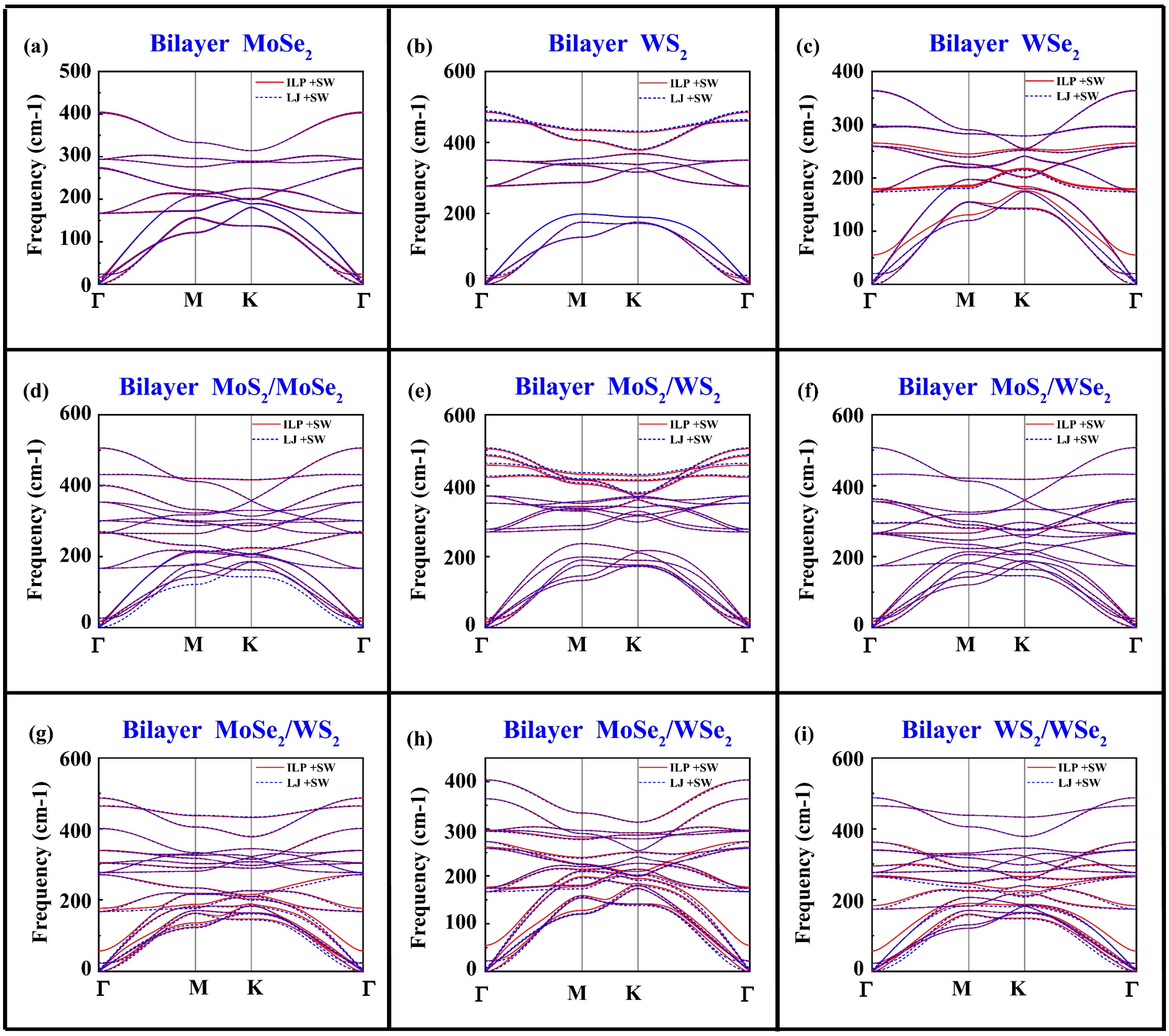

**Figure S11.** Phonon spectra calculated using the ILP (red lines) and the LJ potential (dashed blue lines) with the SW intralayer potential[11] for bilayer homo- (a-c) and heterojunctions (d-i).

In the above calculations, the supercells used for the bulk and bilayer configurations contain 25×25×6 (a total of 45,000 atoms) and 25×25×1 (a total of 7,500 atoms) unit cells, respectively. In both cases, 201 reciprocal space k-points were used to calculate and plot each branch of the phonon spectrum. We note that the choice of interlayer potential affects mostly the low energy phonon dispersion (around the $\Gamma$-point) that corresponds to the interlayer phonon modes.